\documentclass[10pt,preprint2]{aastex}
\newcommand{\bm}[1]{\mbox{\boldmath $#1$}}

\newcommand{\ave}[1]{\left\langle#1\right\rangle}

\newcommand{\simless}{\mathbin{\lower 3pt\hbox
   {$\rlap{\raise 5pt\hbox{$\char'074$}}\mathchar"7218$}}} 
\newcommand{\simgreat}{\mathbin{\lower 3pt\hbox
   {$\rlap{\raise 5pt\hbox{$\char'076$}}\mathchar"7218$}}} 

\newcommand{\de}{$^{\circ}$}
\newcommand{\be}{\begin{equation}}
\newcommand{\ee}{\end{equation}}

\begin{document}

\title{Interstellar Scintillation Observations of 146 Extragalactic
Radio Sources}

\author{\sc Barney J.\ Rickett\altaffilmark{1}}
\affil{Department of Electrical \& Computer Engineering \\
University of California San Diego \\
La Jolla, CA 92093-0407}
\author{\sc T.\ J.\ W.\ Lazio\altaffilmark{2}}
\affil{Remote Sensing Division \\
Naval Research Laboratory\\
Washington DC 20375}


\author{\sc Frank D.\ Ghigo\altaffilmark{4} }
\affil{National Radio Astronomy Laboratory \\
Green Bank\\
WV 24944}

\altaffiltext{1}{e-mail: bjrickett@ucsd.edu}
\altaffiltext{2}{e-mail: Joseph.Lazio@nrl.navy.mil}
\altaffiltext{4}{e-mail: fghigo@nrao.edu}

\begin{abstract}

From 1979--1996 the Green Bank Interferometer 
was used by the Naval Research Laboratory
to monitor the flux density from 146 compact radio sources
at frequencies near 2 and 8~GHz.
We filter the ``light curves'' to separate intrinsic variations
on times of a year or more from more rapid interstellar scintilation
(ISS) on times of 5--50 d.  Whereas the intrinsic variation
at 2~GHz is similar to that at 8~GHz (though diminished in amplitude),
the ISS variation is much stronger at 2 than at 8~GHz.
We characterize the ISS variation by an rms amplitude and a timescale
and examine the statistics of these parameters for the 121 sources
with significant ISS at 2~GHz.  We model the scintillations
using the NE2001 Galactic electron model
assuming the sources are brightness-limited.   

We find the observed rms amplitude to be in general agreement with the model,
provided that the compact components of the sources have about 50\% of their 
flux density in a component with maximum brightness 
temperatures $10^{11}$--$10^{12}$K.   Thus 
our results are consistent with cm-wavelength
VLBI studies of compact AGNs, in that the maximum brightness 
temperatures found are consistent with  the inverse synchrotron limit
at $3 \times 10^{11}$ K, boosted in jet configurations
by Doppler factors up to about 20.
The average of the observed 2~GHz ISS timescales is in reasonable 
agreement with the model at Galactic latitudes above about 10\de.
At lower latitudes the observed timescales are too fast, suggesting that
the transverse plasma velocity increases more than expected 
beyond about 1 kpc.

\end{abstract}

\keywords{ISM: general --- scattering --- plasmas 
--- galaxies: active --- galaxies: quasars --- galaxies: BL Lacertae objects}

\today

\section{INTRODUCTION}
\label{sec:intro}

Variations in the radio flux density of extra-galactic sources have
been observed for many decades; see early observations by
Dent~(1965). Via the classic light travel time argument, variability
timescales have been used to infer the physical size of the radio
emitting regions, and the associated apparent brightness temperature.
Kellerman \& Pauliny-Toth~(1968) reviewed the early work on centimeter
wavelength variations in terms of radiation from (non-relativistic)
expanding synchrotron sources.  In~1981 the same authors reviewed the
topic in the light of very long baseline interferometry (VLBI)
observations of apparent superluminal expansion. Since that time a
standard picture has emerged of relativistic jets from active galactic
nuclei (AGNs), which emit Doppler-boosted radio beams.  If the beam
points close to the Earth, these are seen as very compact sources with
flat radio spectra. The relativistic analysis of such beams
successfully explains the sporadic brightenings which emerge from a
central core and travel out ``superluminally.''  In some models the
brightenings are thought to be shocks and in others physical
structures moving with the jet.

This now canonical jet model has typical Doppler factors in the range
1--20. Flux variations are converted to a ``variability brightness
temperature,'' calculated by the light travel time argument; this
increases as the cube of the Doppler factor over the $\sim 10^{12}$~K
limit due to inverse Compton scattering (e.g., Blandford \&
K\"{o}nigl~1979) or over the $\sim 3 \times 10^{11}$~K 
equi-partition limit as proposed by Readhead (1994). 
Typical year-long flux variations at centimeter
wavelengths give apparent brightness $10^{14}$--$10^{15}$~K, which can
be accomodated within the expected range of Doppler factors.  However,
variations of flux at frequencies below about~1~GHz proved harder to
fit in to the jet model. For a source of a given flux density and
angular size, the associated brightness temperature scales as the
inverse square of the frequency.  Thus the observation of substantial
variations over a year at a frequency of~0.4~GHz (e.g., Fanti et
al.~1981) pushed the apparent brightness to over $10^{16}$K and
stretched the limits of the relativistic beaming models, by requiring
higher Doppler factors to explain the low-frequency variations (LFV)
than to explain the superluminal expansions.

While the discovery of pulsars confirmed that radio sources can vary
intrinsically on very short timescales, their flux variations over
minutes to hours were quickly found to be due to scintillation caused
by inhomogenities in the interstellar medium (ISM).  As reviewed by
Rickett~(1977) these phenomena were found to be consistent with a very
wide range of length scales in the ionized ISM, which causes strong
scintillations (rms $\simgreat$ mean).  Dennison \& Condon~(1981)
looked for and set upper limits to such scintillations from a variety
of compact extragalactic sources including several LFV sources. The
absence of interstellar scintillation (ISS) must be due to quenching
by the angular extent of the emitting regions, from which they
inferred lower limits to the angular size, which were larger than
those inferred from an intrinsic explanation of LFV.

This paradox was resolved by the recognition of a further regime of
ISS. At meter wavelengths pulsars show flux variations over days to
months as well as over minutes.  Sieber~(1982) demonstrated that the
slower variations were also a propagation phenomenon, which Rickett,
Coles \& Bourgois~(1984) identified as predicted by strong
scintillation theory.  This is now referred to as refractive
interstellar scintillation (RISS) to distinguish it from the
diffractive scintillation (DISS) over minutes to hours. Since the
angular diameter that quenches RISS is substantially larger than for
DISS, these authors proposed that RISS could also account for
\hbox{LFV}. The suggestion that LFV was due to irregular interstellar
propagation had earlier been made by Shapirovskaya~(1978), who
proposed that refraction in discrete large scale ionized structures
was responsible.  Several authors (e.g., Rickett~1986; Spangler et
al.~1993) have since demonstrated that RISS does indeed provide a
satisfactory explanation for LFV, based on a power-law model for the
spectrum in the interstellar plasma density.

A further variability phenomenon at centimeter wavelengths was
discovered by Heeschen~(1984). In careful daily measurements he found
a ``flickering'' at a level of a few percent over~1--2~days in the
radio flux from many flat spectrum sources. The sources which showed
the greatest amplitudes were followed up in extensive monitoring
campaigns, particularly, by the Bonn group. Quirrenbach et al.\ (1992)
presented the results of daily monitoring of a complete sample of flat
spectrum sources at~6 and~11~cm. They found a low level of what they
called intraday variability (IDV) in most of the sources observed,
which also had compact VLBI structure.  They concluded that this low
level of variabilty seen in almost all such sources was
\hbox{RISS}. However, in addition they argued that about 25\% of flat
spectrum sources also showed intrinsic variations that can reach
amplitudes as high as 20\%.  An intrinsic variation of~0.2~Jy on a
timescale of~12~hr implies a variability brightness over~$10^{19}$~K,
implying a Doppler factor over~200.  Thus for those sources said to be
showing intrinsic IDV, flux variations were interpreted in terms
extreme brightness emission that strains the limits of the
relativistic jet models.

In fact there should be a continuum of ISS behavior between LFV and
\hbox{IDV}. The wavelength dependence is determined by a combination
of the steep increase with wavelength for the characteristic scale of
RISS and the wavelength scaling of effective source diameter.  For
most extragalactic sources their angular size partially quenches the
scintillations and the effective scintillation amplitude and time
scale are determined by the compact source size, which scales
approximately as wavelength (for brightness temperature limited
emission).  In an analysis of the flux variations of quasar 1741$-$038
over a wide range of frequencies, Hjellming \& Narayan~(1986)
identified ISS at a frequency of~1.49~GHz and slower intrinsic
variations at~15 and~22~GHz.  Dennison et al.~(1987) showed that RISS
was the dominant cause of the 2.7~GHz variation of two sources at low
Galactic latitude, which were in the Green Bank Interferometer
monitoring program.
  
A detailed analysis of one of the large amplitude IDV sources
(0917$+$624) by Rickett et al.~(1995) demonstrated that ISS could
successfully account for the IDV with a source brightness of about $6
\times 10^{12}$~K, providing that the scale height of the scattering
medium was reduced to~200~pc, from the 500~pc scale height of the
Taylor \& Cordes~(1993) model for the Galactic distribution of
electrons.  This was one of the sources for which 20\% intrinsic IDV had
been argued earlier.

An epoch of extremely rapid variability was reported for quasar
0405$-$385 by Kedziora-Chudczer et al.~(1997). They observed large
amplitude variations on timescales as short as one hour at~4.8
and~8.4~GHz.  They considered the implied variability brightness
$10^{21}$K to be unreasonable and argued for an ISS explanation. 
Rickett et al. (2002) reported detailed analysis of the IDV in the 
polarization of this source, which they found to be consistent
with a maximum brightness temperature of $2 \times 10^{13}$K.

Two other quasars have also shown substantial variation over times as short
as an hour (intra-hour variation -- IHV), namely J1819+385
(Dennett-Thorpe \& de Bruyn, 2000, 2002 \& 2003) and 
B1257-326 (Bignall et al.\ 2003). The measurements of a
time delay in the light curves recorded on trans-continental
baselines and the observation of annual modulations in their characteristic
timescales give irrefutable evidence for scintillation as the cause 
of the variations for these two quasars.  It is also important to note that
the three IHV sources mentioned here require that the distance to the
scattering medium be only tens of parsecs rather than the 500-2000~pc
distances involved in most pulsar ISS. A consensus 
has finally emerged that the IDV (and IHV) phenomena are due to ISS.  
The strongest argument in support of intrinsic IDV comes
for one particular quasar (0716$+$714), for which 6~cm IDV was
observed to have some correlation in behavior with optical IDV
observed simultaneously (Wagner et al.~1996).  However, the argument
is weakened by the fact that the optical and radio variations were not
correlated in detail and there was significant decorrelation across the radio
wavelengths observed (see observations by Quirrenbach et al., 2000).

Variations over months and years have been well established at shorter 
wavelengths, as for example reported by Stevens et al.\ (1994)
who monitored a group of blazars for several years at 22 to 375~GHz.
They found reasonable agreement with shock models for the many bursts
they observed. They also observed faster variations at lower levels,
but the brightness temperature deduced by assuming intrinsic variations 
are not extreme at their high frequencies.

Most variability studies are based on relatively small samples 
of sources, often observed at different telescopes using
different observing methodologies.  Lazio et al.\ (2001, henceforth
L01) presented the light curves and structure functions for a sample
of~146 compact radio sources monitored at the Green Bank
interferometer (GBI) over long durations (3--15~yr) with frequent flux
density measurements and a common observing methodology. They
concluded that the rms variation in flux density at both frequencies
was dominated by slow large amplitude changes, due mostly to intrinsic
variability on timescales from a few months to a few years. They also
found more rapid variations on timescales of~5--50~days, which they
characterized as being \hbox{ISS}.

This paper reports further analysis of the GBI flux monitoring
program.  We shall show that the variations at~2.25~GHz do indeed form
a continuum with the ISS phenomena in LFV and \hbox{IDV}.  The plan of
this paper is the following.  In \S\ref{sec:observe} we summarize the
GBI monitoring program.  In \S\ref{sec:data.analysis} we describe our
analysis. In \S\ref{sec:model} we describe a model for the ISS,
which we compare with the observations in \S\ref{sec:compare}. in
\S\ref{sec:individual} we discuss the results for a few sources
of particular interest and we give our conclusions in 
\S\ref{sec:conclusion}.

\section{Observations}
\label{sec:observe}

Flux density observations with the GBI were made nearly daily
from~1979--1987 on 46 radio sources, as described by Fiedler et al.\
(1987b, henceforth F87) and Waltman et al.\ (1991, henceforth W91).
The sample of sources was expanded to~146 and the frequency of
observations reduced to once every 2~days during~1988--1996 (L01).
Here we summarize the relevant details of the Navy-NRAO GBI monitoring
program; for complete details see F87, W91, and~L01.

Sources were selected for monitoring on the basis of their known
variability in flux density and/or compact structure.  Flux densities
were obtained at two frequencies in S-band and X-band.  Until~1989
August the frequencies were 2.7~GHz (S band) and 8.1~GHz (X band)
after which they were changed to~2.25~GHz (S band) and 8.3~GHz (X
band) when new cryogenic receivers were installed.

The observations were made on a baseline of~2.4~km and the correlated
flux density estimated from~15~min.\ integrations of~35~MHz bandwidth
summing right and left circular polarizations.  Each source was
normally observed near the same hour angle each day, but on occasion a
single source was observed at multiple hour angles on the same day.
The flux density scale was set by observations of 3C~286.

The typical rms errors in flux density (L01) are, before~1989 August:
\be
\sigma_2^2 = 0.037^2 +(0.014 S_2)^2
\; \; \; \; \sigma_8^2 = 0.057^2 +(0.049 S_8)^2
\label{eq:sigma1}
\ee
and after~1989 August:
\be
\sigma_2^2 = 0.0037^2 +(0.015 S_2)^2
\; \; \; \; \sigma_8^2 = 0.0057^2 +(0.05 S_8)^2
\label{eq:sigma2}
\ee
Here $S_2$ and~$S_8$ are the flux densities in the S- and X band,
respectively.  

Attempts to estimate the magnitude of systematic errors suggest that
these may approach 10\% at S-band and 20\% at X-band.  There is also a
systematic ripple, up to approximately 5\% peak-peak in a few
sources, introduced by the calibration observations of {3C286}.  
Its origin is unknown ({F87}), but its 1-yr timescale suggests 
a seasonally-induced instrumental effect.

Clearly discrepant data were edited out of the published time series
by removing those data that deviated by more than 10$\sigma$ at X band
or~15$\sigma$ at S band from a running boxcar mean.
Nonetheless, as can be seen in many of the light curves (L01,
Figure~8), there are occasional obvious one-day deviations remaining in
the light curves.  We do not regard these as valid rapid changes in
flux density and we removed them from the time series by a two-step
editting process. In each step each data point was compared with a
smooth model value obtained from a parabolic fit to a 365-day
subinterval centered on that sample.  If it deviated by more than
3.5$\sigma$ it was rejected; here $\sigma$ is the rms deviation
between the data and the model over that subinterval.  The rejected
points have some influence on the $\sigma$ and so in a second step the
procedure was repeated, rejecting a few more points. With two editting passes the
numbers of points rejected from each source in each band 
were typically 10--30 in 1988--1996 but none in 1979--1987.  
A few of the rejected points were from the same day 
at S and X band, which would be due to a common failure 
in the receiver system or due to a calibration error common to both bands.

We analyzed the data in two sets --- 46 sources
during 1979--1987 at 2.7 GHz and 8.1 GHz  and 146 sources
during 1988--1996 at 2.25 and 8.3 GHz.
As noted above the second set started with about 18 months at
2.7 GHz. While the frequency change caused little discontinuity
for most sources with relatively flat spectra,
there were some for which the frequency change
caused a substantial discontinuity in flux density. 
In such cases we scaled their 2.7 GHz flux densities
to 2.25 GHz using the mean spectral index between 2.25 
and 8.3 GHz.

After the conclusion of the GBI monitoring program, Fey \&
Charlot~(1997, 2000) quantified the compactness of many of the sources
monitored.  They determined the contribution of the sources'
structures to interferometric group delays in VLBI observations and
characterized the sources by ``structure indices'' ranging from~1
to~4.  A source with a structure index of~1 is relatively compact on
scales of order 1~mas, having 90\% or more of its flux density in a
single component, while a structure index of~4 indicates a relatively
extended source, typically having two or more components of
approximately equal flux density. The structure indices were
listed in Table~I of L01 and are used in our analysis below.  
From models fitted to these and other VLBI observations in S-band
we obtained the flux density and diameter of the
most compact source component.  However, we
re-iterate a caution made in L01.  It is well known that sources may
show structural changes on milliarcsecond scales on timescales of
decades.  Since the structural estimates are computed from
VLBI observations after the end of the GBI monitoring program,
they should be taken only as indicators of a
source's structure during the flux density monitoring.

\begin{figure*}
\includegraphics[width=8cm,angle=-90]{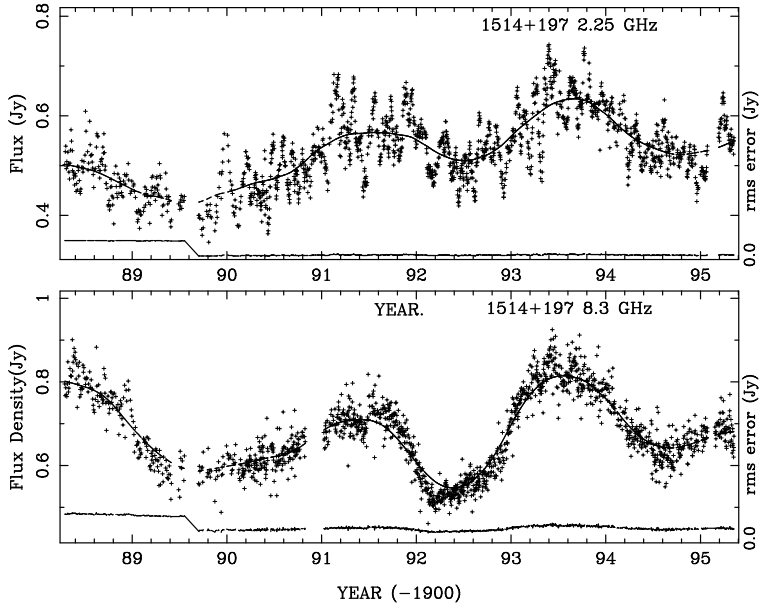}
\figcaption[]{Flux density at 2 and 8 GHz for 1514$+$197.  The solid curve
is the time series in Jansky smoothed over~365~days with a truncated cosine
function.  It is interpreted to be dominated by intrinsic variation in
the synchroton radiation from the source. The short term deviations
from the line are due to \hbox{ISS}.  The trace below each curve gives
the rms error in a single measurement (on the same scale)
relative to the horizontal axis.
\label{f:flux1514} }
\end{figure*}

\section{Data Analysis}
\label{sec:data.analysis}

\subsection{Methodology}\label{sec:method}

\begin{figure*}
\includegraphics[width=8cm,angle=-90]{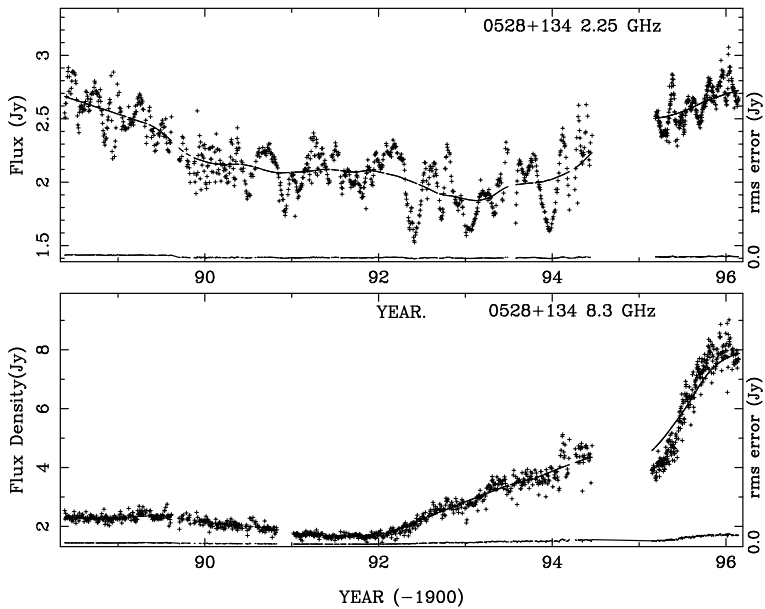}
\figcaption[]{Flux density at~2 and~8~GHz for 0528$+$134 in the same format
as Figure~\ref{f:flux1514}.
\label{f:flux0528} }
\end{figure*}

We motivate our analysis by considering the blazar 
1514$+$197\footnote{Throughout the paper we refer to sources by their 1950
coordinates}.
Figure~\ref{f:flux1514} shows its light curve from~1988.3 to~1995.3.
Immediately evident is that the 2~GHz variations are much more rapid
(5--50~days) than the 8~GHz variations, which encompass only three
main peaks in~7~yrs.  These slow variations also appear at~2~GHz, but
with a reduced amplitude and significant delay compared to those
at~8~GHz, as expected for intrinsic synchrotron bursts.  In contrast,
the fast variations are relatively strong at~2~GHz and much weaker at
8~GHz, which is consistent with the expected decrease in ISS amplitude
with increasing frequency. In most of the sources monitored there 
are 2~GHz variations present on timescales from 5--50 days but at 
somewhat lower amplitudes than for 1514$+$197.
\emph{Consequently, we adopt the hypothesis that the rapid variations 
visible in most of the sources at~2~GHz are dominated by ISS and the 
slower (year-long) variations at both frequencies are intrinsic.}
Figure \ref{f:flux0528} shows a source at a lower Galactic latitude
(-11\de) with similar behaviour but with a longer timescale at 2~GHz
than for 1514$+$197, which is at latitude 56\de.

We separated the fast and slow variations by applying
low-pass and high-pass filters to the light curves.  For each time
series we smoothed the data by a cosine function truncated at its
first zeroes with a full width $T_{\rm sm}$.  This smoothed series
characterizes the slow variations and was subtracted from the raw
series, leaving the difference as the high-pass time series.  In the
figures the smoothed data are shown by the solid lines and the
high-pass time series is simply the deviation of the individual points
from the smooth curve.  We initially set $T_{\rm sm}$ to~365~days at
both 2 and~8~GHz, since typical time-scales for cm-wavelength
variations have been a year or more (e.g., Medd et al.~1972; Hughes,
Aller \& Aller~1992).  
For a few sources there were
substantial slow variations which were nevertheless 
faster than the 365-day smoothed curve.  
For these sources we adopted a shorter smoothing time in the
range 75--200~days as listed in Table~\ref{tab:results}.  
Several of these sources are well-known variables 
at cm-wavelengths and the separation of
intrinsic variations from ISS variations is less clear than for
those with intrinsic times-scales of a year or more.

\begin{figure*}[htb]
\begin{tabular}{lr}
\includegraphics[width=6cm]{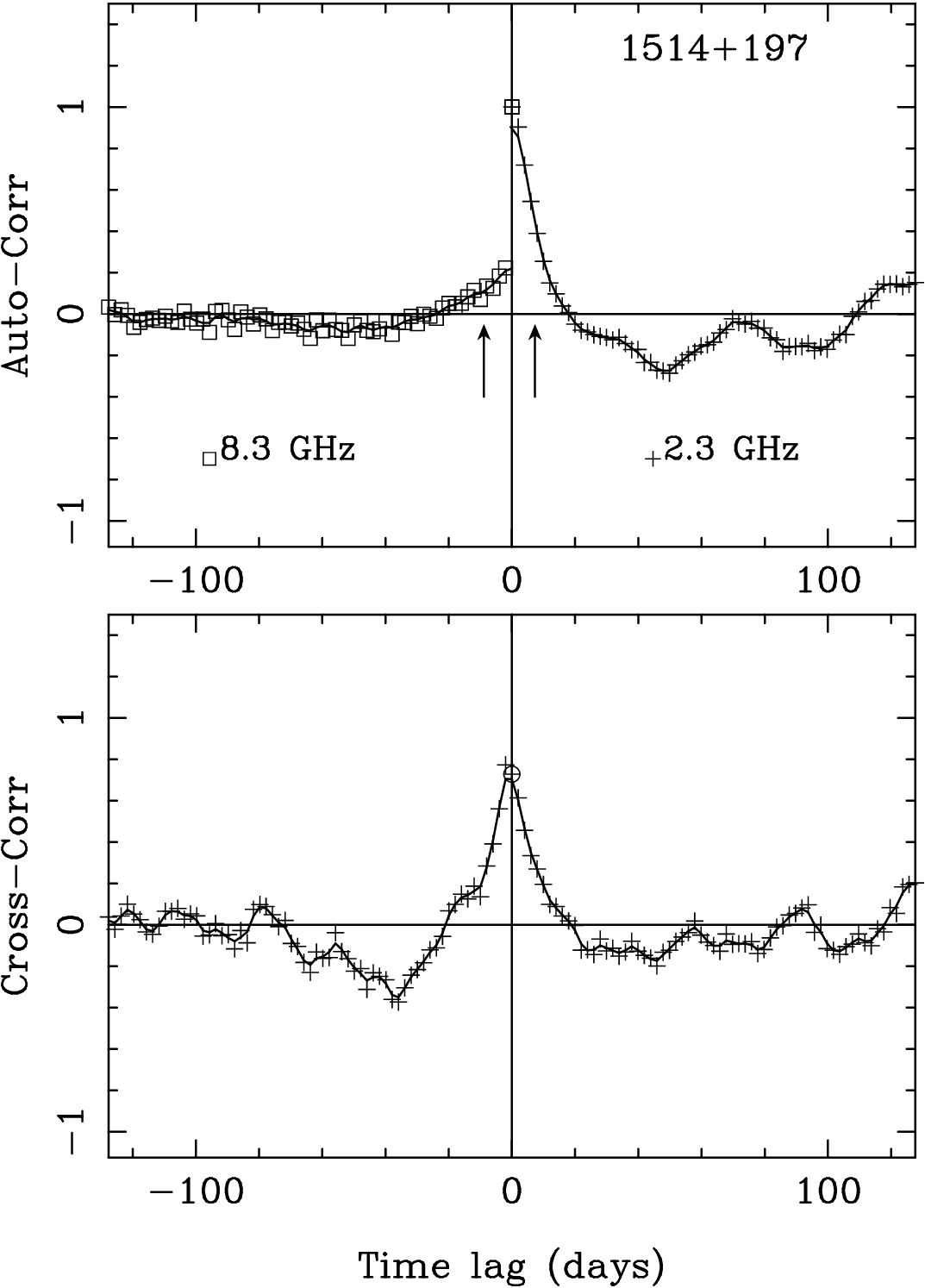} &
\includegraphics[width=6cm]{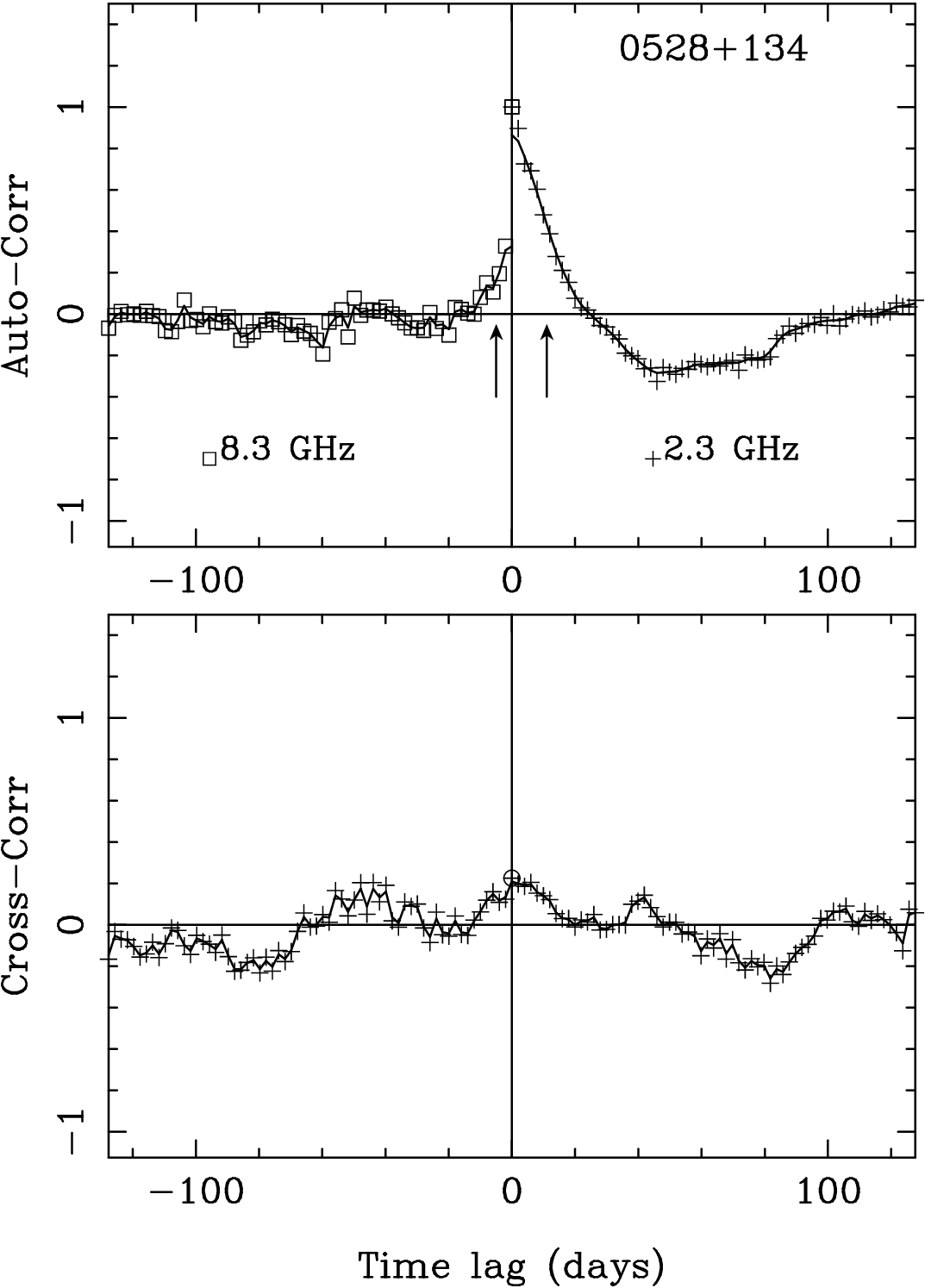}
\end{tabular}
\figcaption[]{Correlation functions for the
high-pass filtered time series for 1514$+$197 (left) and
0528$+$134 (right) at~2 and~8~GHz. 
\textit{Top}: Normalized auto-correlations with the
8~GHz acf plotted against negative time lags.
The overplotted lines are smoothed as described in the text.  The
signal variance is found from the zero lag value of the smoothed curve
before normalization. The timescale estimates are indicated by vertical
arrows.
\textit{Bottom}: The cross correlation of the 2 and~8~GHz time
series, normalized by the square root of the product of the two signal
variances.
\label{f:acf1514} }
\end{figure*}

In order to study the scintillation phenomenon in the short
term variations, we want to normalize them by the mean intrinsic
flux density of each source.  Since many of the sources do 
vary intrinsically, as characterized by the low-pass time series,
we divided the high-pass series by the low-pass series for each source.
We refer to the result as the scintillation (ISS) data,
from which we estimated a timescale and an rms
amplitude.  For each source we
computed the auto-correlation function (acf) at both frequencies and
the cross-correlation function (ccf) between them. Though the data
were sampled at nominal intervals of either 1 or 2 (sidereal) days,
the actual sampling is not exactly regular. Thus we computed the acf
by summing the lagged products into bins at lags of an integer times
the nominal sample interval, normalizing the sum by the actual number
of products in each bin.  Figure~\ref{f:acf1514} shows examples.

In general the acfs have a narrow spike at the origin, due to noise, a
broader component decaying toward zero over times of~5--50~days but
typically oscillating randomly about zero at larger lags; superposed
on this are noise variations that are independent from one lag to the
next.  It is the ``signal'' with a timescale of~5--50~days that
we attribute to \hbox{ISS}.  We reduced the effect of the noise by
smoothing each acf by a Gaussian function with a 3.3~day full width at
half-maximum, but first replaced the zero lag point by the acf at the
first lag (1 or~2~days). The signal variance was then estimated from
this smoothed acf at zero lag. This procedure assumes that the ISS
timescale is several days; any rapid (intra-day) variations will not
be included in such an estimate but will contribute
to the ``noise''.  The rms amplitude is taken as the
square root of this signal variance giving the short-term 
modulation index $m = S_{\rm rms}/\ave{S}$,
since we have already divided by the (varying) source
flux density.  The signal timescale~$\tau$ was found as the lag
where the smoothed acf crosses half of the signal variance. 

This analysis, in principle, allows us to estimate 
$S_{\rm rms}$ to values well below equation~(\ref{eq:sigma2}), 
which applies to a single observation of flux density.  
However, the limit in the equation near $m = 0.015$ is mostly due to
uncertainties in the calibration of flux density, which is subject to
low amplitude errors -- some of them systematic. 
These may be influenced by weather, equipment
changes and other factors that are not well monitored.  Thus, even
though our $m$ estimates are corrected for the influence of noise and
other measurement errors that are independent from one day to the
next, we have little confidence in identifying ISS where $m$ is less
than about 0.01.

In analyzing the same data, L01 characterized the
short-term variabilty by the logarithmic slope of the 
intensity stucture function computed between lags of 2 and 32 days.
They found values between 0 and 1.2, with somewhat
higher structure function slopes at lower Galactic latitudes,
suggesting that variations on times of 32 days or shorter were 
due to ISS.   We have plotted this parameter against our 
modulation index $m_2$, and find them to be highly correlated.
In general a value of 0.67 is expected for the structure
function slope for pure ISS in a Kolmogorov scattering medium that extends
up to the observer (see Coles 1988, Blandford, Narayan 
\& Romani, 1986).  However,
the structure function estimated by
L01 was not corrected for the influence of system noise.  
Since the structure function for 
pure white noise is flat, the slope parameter will be 
significantly reduced when the variance due to ISS 
is comparable to that due to system noise.
We use our $m_2$ estimate to characterize the ISS,
since it is corrected for noise. We return to the question of
the structure function slope in \S\ref{sec:ISM-model}.

Table~\ref{tab:results} tabulates the results
with 1950 coordinates as the source name with Galactic
coordinates and effective length through the interstellar 
plasma (see \S\ref{sec:issmodel}).  We omit
SS433 (1909$+$048) from the Table and include the Galactic 
center source Sgr A* (1742$-$289), but neither are considered 
in the statistical analyses that follow.
We give mean flux densities, modulation index 
and timescale as $m_2$, $m_8$, $\tau_2$ and~$\tau_8$ at~2 and~8~GHz.  
The formal errors in these parameters are typically
10-20\%, but the acf plots sometimes showed large
fluctuations versus time lag. By inspecting each plot 
we assigned an ``ISS confidence level'' (0,1,2) 
at each frequency:  2 is for convincing evidence 
for short-term variations (errors 10-20\%); 
1 is for just credible evidence (errors 20-50\%); 
0 when no credible short-term variations are detected (even though
$m$ and $\tau$ were formally estimated).  This ISS confidence level
is given in brackets following the enries for $m_2$, $m_8$.
While at~2~GHz 121
sources show a credible ISS ``signal'' and a timescale in the range
5--50~days, only 60 sources show credible ISS component at~8~GHz. 
For the sources observed in 1979--87 as well as in 1988--96, 
the parameters were estimated separately from each data set.
We find a strong correlation (85\%) between the two 
independent estimates of $m_2$
and somewhat lower correlation (42\%) between the two estimates
of  $\tau_2$. In Table~\ref{tab:results} 
we list the averages from the two epochs for the 38 sources
common to both data sets, and we omit the 5 sources
in the first but not in the second data set.

The ccfs were computed in the same fashion but normalized by the
square root of the product of the signal variances at the two
frequencies.  If the varying signal were fully correlated between~2
and~8~GHz, the two acfs would have the same shape (though perhaps
different variances) and the ccf would show the same symmetrical shape
peaking near unity at zero lag.  Few of the observed ccfs follow
such ideal behaviours but typically show random-looking variations
over time lags of~10~days or more with superimposed noisy
(day-day) variations.  Several of the sources with large amplitude slow
variations show partial leakage of the intrinsic variation through
the high-pass filter; this appears as a cross-correlation peaking 
at positive lags, which corresponds to 8~GHz variations preceding
those at~2~GHz. 

It was found also that several sources exhibited a low amplitude
spike in the ccf at zero lag.  Since these were 
mostly sources with an ISS confidence level of 0 and 1, 
the spike is not due to very rapid wide-band ISS. 
Of course, the additive noise in the time series
should be entirely independent at the two frequencies. Thus the spike
is an unexpected correlation introduced perhaps in the calibration
process, due to the influence of a temperature change on system gain,
bad weather, or some similar systematic effect, which varies
independently from day to day.  To avoid this
influencing our cross-correlation estimates we smoothed the ccfs
(ignoring the spikes) in the same fashion as for the acfs and recorded
the smoothed ccf at zero lag as a measure of the cross-correlation
coefficient.  In the table we list this zero-lag correlation 
coefficient only for the 36 sources with ISS confidence level
of two at 2~GHz and one or more at 8~GHz. Among these several 
cross correlations are influenced by both ISS and 
relatively rapid intrinsic variations which typically 
shift the correlation peak to positive time lags, as noted above.

\subsection{Overview of Variations}
\label{sec:classify}

We have completed the analysis described above on all of the sources
and have tried to classify their variations, by examining
all of their 2 and~8~GHz light curves (see Figure~8 of L01) and
the correlation plots as in Figure \ref{f:acf1514}. 
In addition to listing the parameters of 
the short-term variations and the ISS confidence levels,
we also list in Table~\ref{tab:results} 
a characterization of the slower variations as
$M_2$ and $M_8$, which give the rms variation 
on a 300-day timescale, normalized by the mean flux density.
(This quantity was obtained from the square root of 
one half of the structure functions of the low-pass
time series at a time lag of 300 days for both 2 and 8 GHz).

We find that there are a several types of light curve, 
that depend on the relative importance of ISS and intrinsic
variations.  For some sources (e.g., 0235$+$164, 
1749$+$096, 2200$+$420) intrinsic variations
dominate, showing a sequence of 
outbursts on times of a few months to~2~yrs, often
delayed and reduced in amplitude at~2~GHz relative to~8~GHz, as
expected for intrinsic synchrotron flares. For some of these with
faster outbursts we needed to reduce the smoothing time $T_{\rm sm}$
(as listed in Table \ref{tab:results}), and
so the separation of ISS from intrinsic variations is 
less clear.  At 2~GHz many sources 
show both the long-term (intrinsic) variations and short-term (ISS) 
variations over 5--50~days.  In a few cases (e.g., 1635$-$035) 
there is a larger overall amplitude variation 
at~2~GHz than at~8~GHz.  The 2~GHz short-term timescale $\tau_2$
was somewhat longer for sources at lower Galactic latitudes
(e.g.\ 0528$+$134, 0653$-$033, 1741$-$038), but timescales of
20 days were also seen in a few sources at mid-latitudes 
(e.g.\ 0537$-$158, 1614$+$051, 1635$-$035).
A few sources show the systematic annual variation (e.g., 
5\% peak-peak for 0952$+$179), which is also seen at a low level
(1.5\% peak-peak) in the flux density calibrators 1328$+$254
and 1328$+$307.

There has already been considerable analysis of the
discrete events known as extreme scattering events (ESEs)
in the GBI data (see F87, Fiedler et al., 1994, L01).  
The complete set of time series in L01 
show that there are occasionally other 
events or evidence of non-stationary variation at 2~GHz.
Examples are 1538$+$149 (see Figure \ref{f:flux1538}) 
which from 1989.9 until 1991.6 showed a
dramatic increase in variation on month-long timescales,
and 1502$+$106 and 2059$+$034, which both showed several large 
amplitude excursions which did not meet the criteria for an ESE.
The fact that these ``events'' were much stronger at 2~GHz than
at 8~GHz suggests an origin in changing interstellar propagation 
conditions, but we do not pursue their origin further.

\section{Modeling Interstellar Scintillations and Intrinsic Source Structure}
\label{sec:model}

We now develop a model for the expected ISS of the sources in the GBI
monitoring program.  There are two aspects to this modeling, the ISS
itself and the intrinsic source structure.  We discuss each in turn,
but put the mathematical details of the ISS in
Appendix~\ref{app:iss}.

\subsection{Interstellar Scintillation Model}
\label{sec:issmodel}

From pulsar observations it is known that the ionized ISM has a
disc-like Galactic distribution.  The electron density microstructure
is well-modeled by a power-law power spectrum with a Kolmogorov
spectral index on many lines of sight (Armstrong, Rickett, \&
Spangler~1995).  Taylor \& Cordes~(1993, hereafter TC93) developed a
detailed model for the electron distribution in the ISM, which
has been revised by Cordes and Lazio~(2005, hereafter CL05).
Both papers give an explicit model for the local mean electron
density and
the strength of its small scale variability as characterized by the
strength parameter~$C_N^2$.  This parameter sets the level of the
density spectrum, $P_{3N} = C_N^2 \kappa^{-11/3}$, at
three-dimensional wavenumber~$\kappa$.  In this expression the
inhomogeneities are described as isotropic, since along a given 
line of sight any anisotropies in the density structures are expected to be
randomly oriented, so that the cumulated effect is approximately
isotropic in the transverse plane.  A measure of the cumulated
scattering expected for a particular radio source is given by the
scattering measure defined as the line of sight integral ${\rm SM} =
\int C_N^2 dl$.  Though the spectral index of the density power
spectrum is near~$-11/3$ for many lines of sight, $C_N^2$ and SM 
can vary by orders of magnitude over angular scales of a few degrees.

There are several observable scattering effects, of which we are
concerned with the variability in apparent radio flux density
(scintillation).  Scintillation is classified as weak or strong
according to the scintillation index ($m_{\rm pt}=$ rms/mean flux
density) \emph{for a point source}, 
which depends on the observing frequency as well
as on the scattering measure.  If $m_{\rm pt} \ge 1$, the
scintillations are strong; if $m_{pt} \ll 1$, the scintillations are
weak. The inverse-frequency dependence of the refractive index in a
plasma leads to a decrease in the strength of scattering with
increasing frequency.

Interstellar scintillation was first observed in pulsars at relatively
low radio frequencies (0.1--1~GHz) and was mostly in the strong regime,
in which there are two branches of the phenomenon, diffractive and
refractive scintillation (DISS and RISS as already referred to). The
extremely small diameters of the emitting regions in pulsars ensures
that they scintillate essentially as point sources showing both
branches. See reviews by Rickett~(1990) and Narayan~(1992) for
discussions of the physics of DISS and \hbox{RISS}.  Given the TC93
distribution Walker~(1998, 2001) calculated the transition frequency
separating weak from strong scattering for extragalactic sources; he
found values ranging from about~6~GHz toward the Galactic poles to
about~2~GHz for directions toward the inner Galactic plane.

In contrast, extragalactic sources are observed typically at higher
frequencies (2--15~GHz) and in addition their angular extent is
much larger than that of pulsars.  As a result extragalactic
sources are often observed under weak ISS conditions; however, ISS is
usually not detectable, because their angular extent is usually
sufficient to ``quench'' the scintillations, in the same way that the
angular extent of planets quenches their optical twinkling in the Earth's
atmosphere.  These effects explain the rarity of ISS observed for
extragalactic radio sources.  Low-frequency variability below~1~GHz
(\S\ref{sec:intro}) is due to strong RISS of compact radio sources
(e.g.\ Rickett, 1986) while IDV is weak ISS (WISS). 


The spatial character of the scintillations in a transverse plane can
be characterized by the auto-correlation of the intensity (fourth moment
of the field) or from its two-dimensional Fourier transform versus
transverse wavenumber (the intensity spectrum).  From the theory of
wave propagation in random media (e.g., Ishimaru~1978), we can obtain
a theoretical expression for the intensity second moment, but its
evaluation is difficult under the most general conditions.  However,
the expansion for low wavenumbers leads to the solution for both weak
ISS (WISS) and for RISS in strong scattering; the DISS solution is
obtained in the high wavenumber limit (see Coles et al.~1987, hereafter
CFRC). When the source has a significant angular extent the fine-scale
(high wavenumber) fluctuations of the pattern are smeared out by the
effects of overlapping patterns from neighboring source components.
This explains why no DISS has been observed from any extragalactic
sources (Dennison \& Condon~1981).  It also means that we only need the low-wavenumber expansion
to describe both WISS and \hbox{RISS}.  In this limit the intensity
spectrum is given by a line of sight integral of the product of the
local density spectrum, the Fresnel filter that suppresses the power
at low wavenumbers and a low-pass cut-off due to the visibility of the
source (CFRC, equation~11).

Rickett et al.~(1995) used this method, to interpret the ISS from IDV
source 0917$+$624 observed at frequencies both above and below the
weak/strong transition frequency (see also the analysis of IDV from
quasar 0405$-$385, Rickett et al.~2002).  Here we follow the same
method but generalize the theory to an arbitrary distribution of
scattering along the line of sight.  In Appendix~\ref{app:iss} we give details
of the appropriate expression for the spatial correlation function of
intensity as an integral over the distance~$z$ from the Earth
including $C_N^2(z)$ from the TC93 or CL05 models.

The equations as given in Appendix~\ref{app:iss} yield the 
scintillation index $m_{\rm iss}$ and spatial scale $s_{\rm iss}$,
under an assumed form for the source visibility function
as described in the next section.
We convert the spatial scale to a temporal scale by dividing by an assumed
velocity for the scintillation pattern. There are a few examples of
IDV in which the relevant velocity has been shown to be consistent
with the speed of the Earth relative to the local standard of rest
(LSR).  Consequently we used the annual average of this LSR speed
projected transverse to each line of sight as the model velocity
(~20--30~km~s$^{-1}$),  and so obtained model time-scales
$\tau_{\rm iss}$ for each source.  
As Appendix~\ref{app:iss} shows, the contribution of the scattering
medium to the ISS is controlled by the effective scattering
length~$L_{\rm eff}$ and the scattering angle~$\theta_s$.
In developing the necessary code for the calculations we made
use of the subroutines described by TC93 and CL05.

\subsection{Intrinsic Source Structure Model}
\label{sec:intrinsicmodel}

\begin{figure}[htb]
\begin{tabular}{l}
\includegraphics[width=5cm,angle=-90]{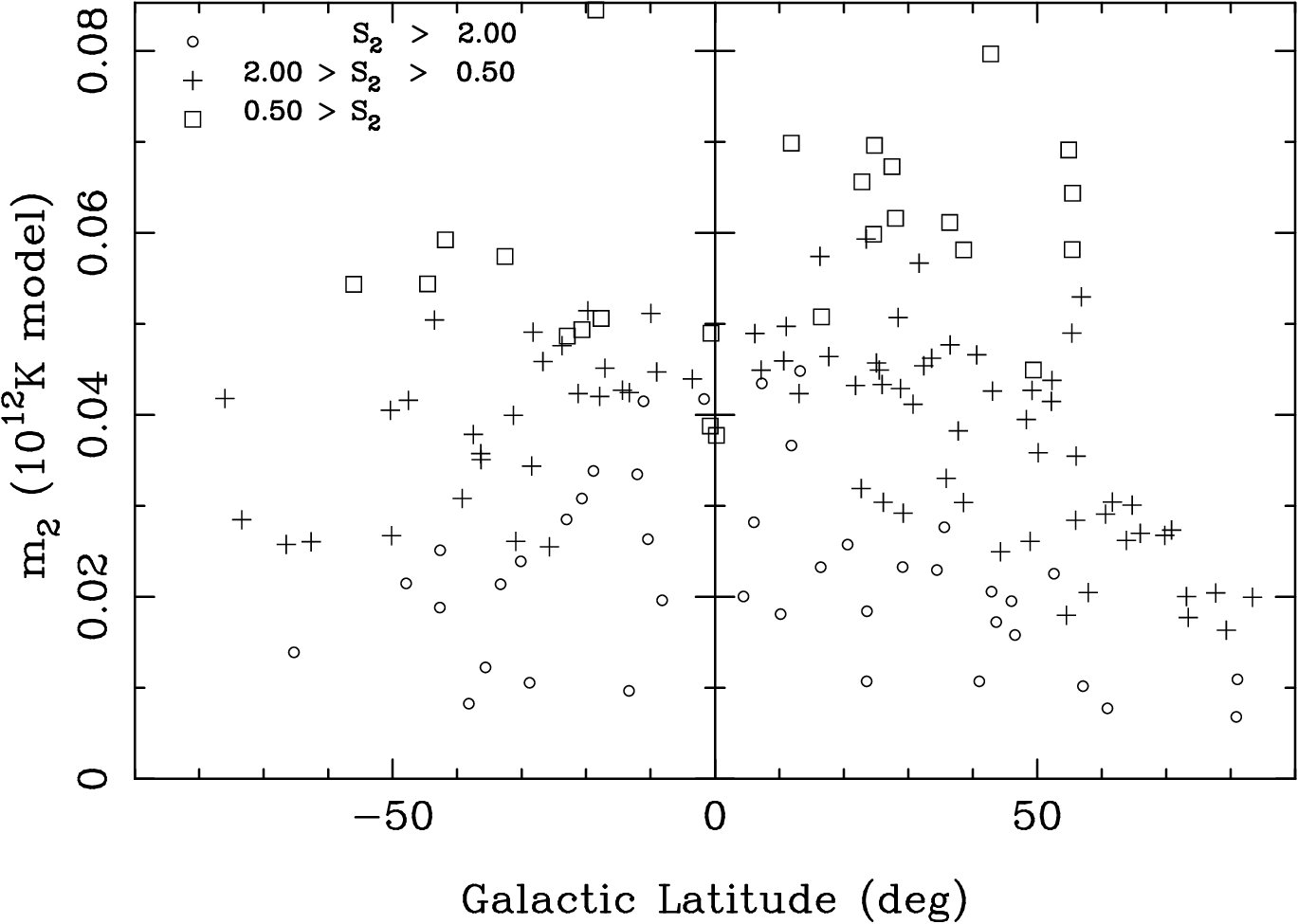} \\
\includegraphics[width=5cm,angle=-90]{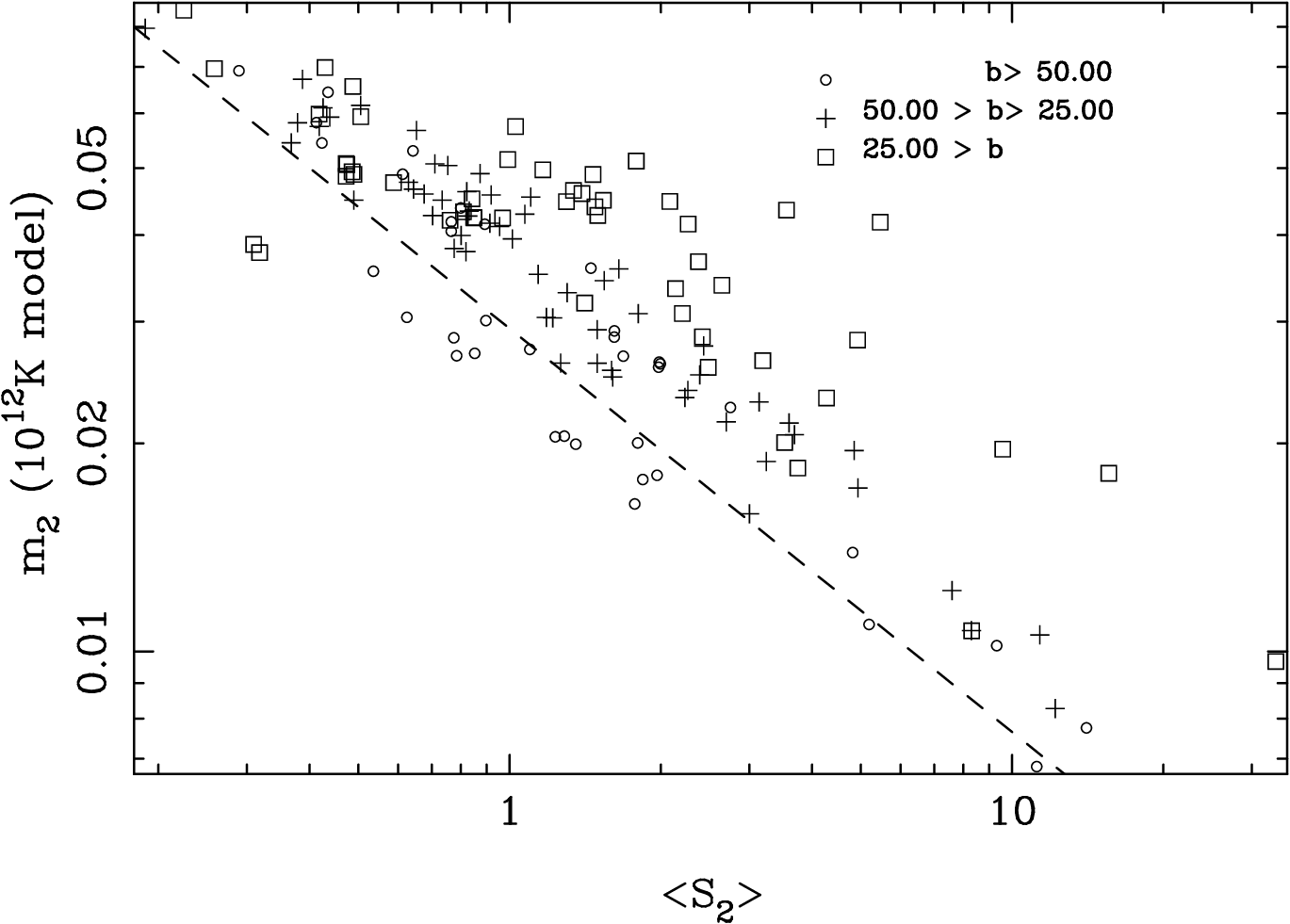} 
\end{tabular}
\figcaption[]{Model scintillation index at 2.25~GHz computed under 
the CL05 model for the Galactic electrons,
assuming that each source has 50\% of its flux
density in a Gaussian core of~$10^{12}$~K brightness.
Model scintillation index versus
\textit{Left}:  Galactic latitude and \textit{Right}:  mean flux density
with dashed line $m_{\rm iss} \propto \ave{S}^{-7/12}$.
\label{f:mt.theor_b} }
\end{figure}

Having specified a model for the scattering medium, we must also
specify a model for the visibility functions for the sources in the
survey.  Readhead~(1994) has argued that the distribution 
of brightness temperatures from relativistically beamed
jets of~5~GHz-selected radio sources ranges from 
$\sim 3\times 10^{10}$ to $3\times 10^{12}$~\hbox{K}. 
Accordingly, we adopt a simple model in which the most compact component 
of each source has a circular Gaussian brightness distribution 
in which the peak brightness is limited to some maximum
due, for example, to synchrotron self absorption or
inverse Compton losses.  Brightness temperatures
up to about~20 times greater than the equi-partition value
of $3\times 10^{11}$~\hbox{K} are seen in VLBI images of 
many radio-loud active galactic nuclei.  
These are well described by relativistic
beaming from narrow jets with Doppler factors up to about~20. Thus we
examine three source models with peak brightness temperatures
of~$10^{11}$, $10^{12}$, and~$10^{13}$~\hbox{K}.  For each 
of the 146 sources, we assumed 50\% of their flux density
to come from a Gaussian brightness core of these three temperatures.
Hence we obtained three diameters (full width at half maximum -- FWHM) using
\be
\theta_{\rm FWHM} = (1.1/f_{\rm GHz}) \sqrt{S_{\rm Jy}/T_{\rm b,12}}
\; \mbox{ mas }
\label{eq:FWHM}
\ee
where the core flux density is $S_{\rm Jy}$,
the peak brightness temperature is $T_{\rm b,12}$ 
in units of $10^{12}$K and $\theta_{\rm FWHM}$ is in mas.

Figure~\ref{f:mt.theor_b} shows
the computed $m_{\rm iss}$ at 2.25~GHz for the $10^{12}$~K source models
under the CL05 model for the electron distribution
versus Galactic latitude and total mean source flux
density~$\ave{S}$. There is only a weak trend for increased
$m_{\rm iss}$ at low latitudes; this illustrates its dependence
on properties of both the ISM and the source.
The right panel shows a lower envelope to the 
(inverse) relation of~$m$ to~$\ave{S}$.  It is approximated
by the dashed line,  whose slope is from CFRC
(equation~13), which with $\theta_{so} \propto \sqrt{\ave{S}}$ gives 
$m_{\rm iss} \propto \ave{S}^{-7/12}$.  

\section{2~GHz Observations Compared with ISS Model}
\label{sec:compare}

In this section we confront the model with the observations.  
We begin with global
comparisons of the observations with ISS predictions from the model. 
Next we compare the fast variations to the slow variations,
since both the fast ISS and slow intrinsic variations are
indicators of compact structure.
Finally, in a limited number of cases where the ratio of the 
ISS signal to the noise is sufficiently strong, we perform two further
analyses.  First we do a power spectrum analysis and compare with theory.
Second we epoch-average the timescale separately for each of the 12 months
in the year, in a search for an annual modulation in the ISS timscale due
to the changing Earth velocity as it orbits the Sun relative to the 
scattering medium. 

\subsection{A Global Comparison for Scintillation Index}
\label{sec:global.m}

\begin{figure*}
\begin{tabular}{lr}
\includegraphics[width=5.5cm,angle=-90]{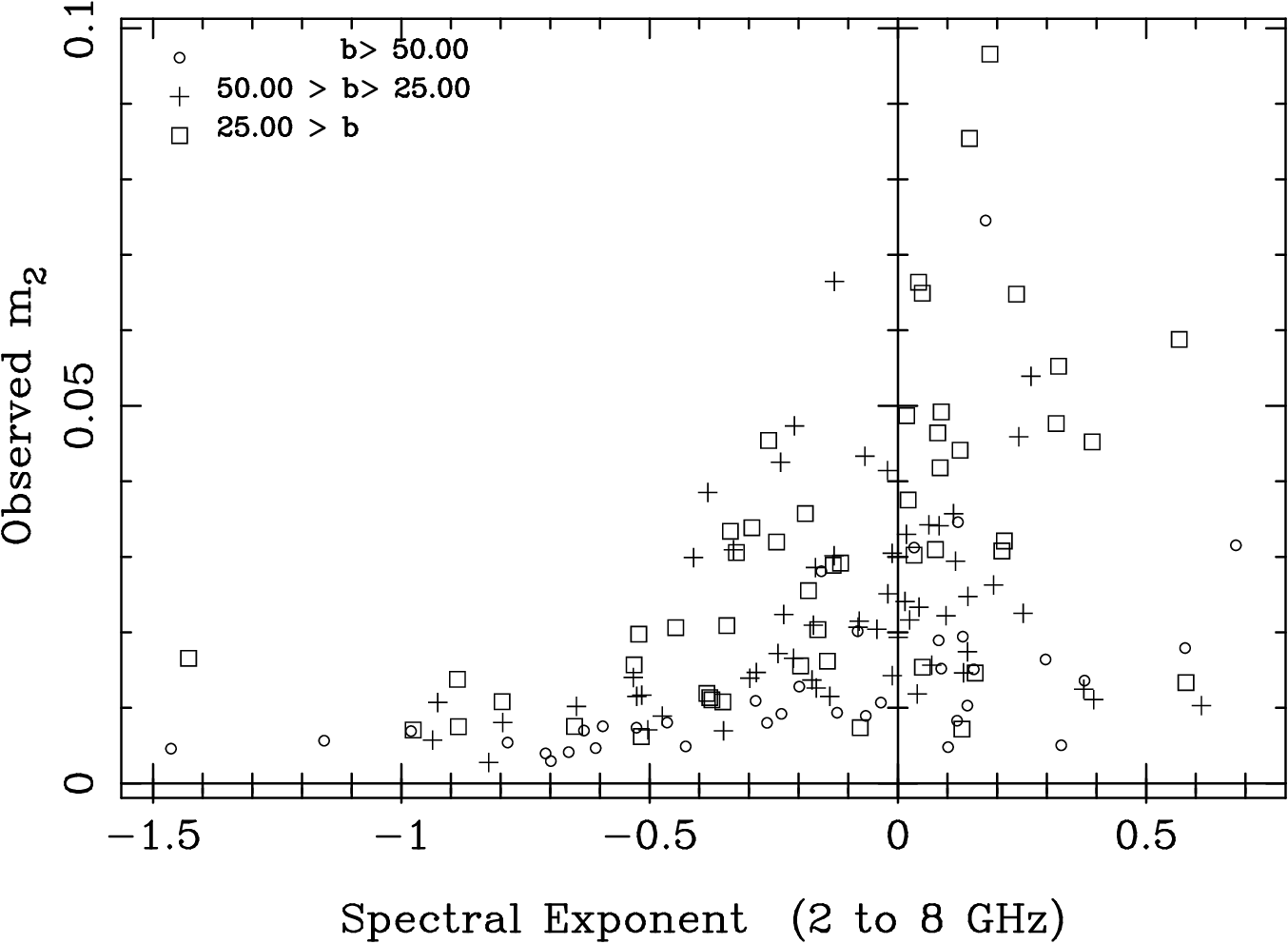} &
\includegraphics[width=5.5cm,angle=-90]{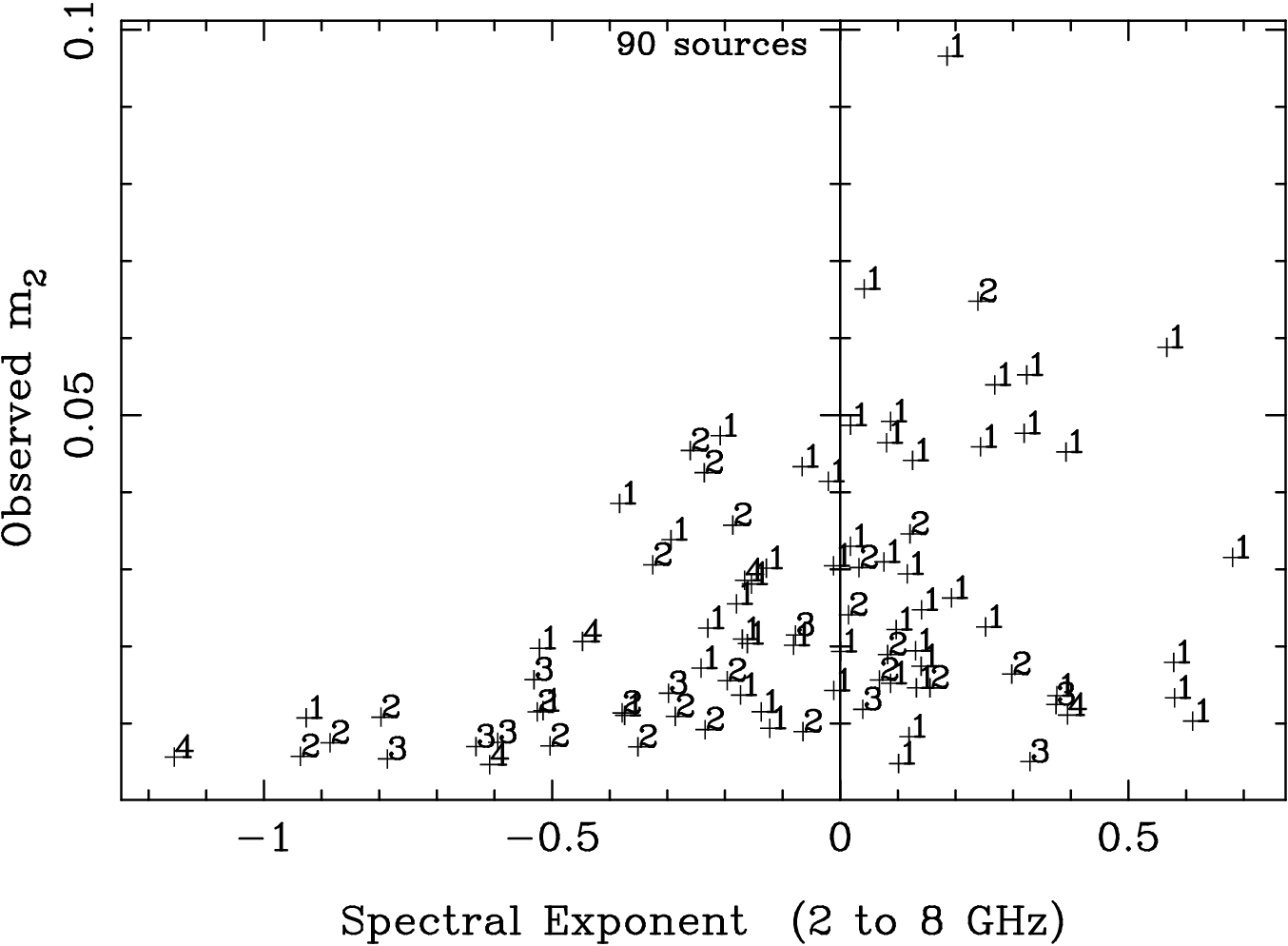}
\end{tabular}
\figcaption[]{Scintillation index at~2~GHz against mean 
spectral index between~2 and~8~GHz.  
\textit{Left}:  The different symbols flag ranges of the magnitude of the
Galactic latitude.
\textit{Right}: Subset of sources with structure index 1--4,
as an indicator of increasing source size. 
\label{f:m_alpha} }
\end{figure*}

Since the short-term variations are much stronger at 2 than at 8~GHz,
we compare the modulation indices~$m_2$ and timescales~$\tau_2$
with the ISS models. In making the comparison, we must consider
the two major influences, which are the source structure and the
distribution of the scattering plasma, for which we use the CL05 model.   
With our simplistic source model (i.e., $10^{12}$~K component with
50\% of the flux density), the results are only meaningful in a
statistical sense. In addition the CL05 model only 
approximates the Galactic distribution of the warm
ionized medium and substantial deviations are to be expected on
particular lines of sight. In many of the figures that follow we
plot the results against a parameter of the scattering 
model and flag the points according to some measure of
the source model, or vice-versa.

Figure~\ref{f:m_alpha} plots the observed $m_2$ against 
mean spectral index ($\alpha$) between 2 and 8~GHz
($S\propto \nu^{\alpha}$), which is also listed in Table 
\ref{tab:results}.  We see that for
relatively flat spectrum sources (say $\alpha > -0.4$) 
there is a much higher
fraction of sources with scintillation indices above 1\%, 
than for steeper spectrum sources, as expected if
steeper spectrum sources have larger diameters.  
In the left panel the points are flagged by the magnitude of
the Galactic latitude ($|b|$), showing the expected trend 
for greater ISS at lower latitudes.   For the steepest
spectra ($\alpha < -0.6$) there is little or no significant
scintillation, and so in the analysis that follows we 
exclude the 17 steep spectrum sources ($\alpha < -0.6$),
which include the flux density calibrators.

\begin{figure*}[htb]
\begin{tabular}{lr}
\includegraphics[width=5.5cm,angle=-90]{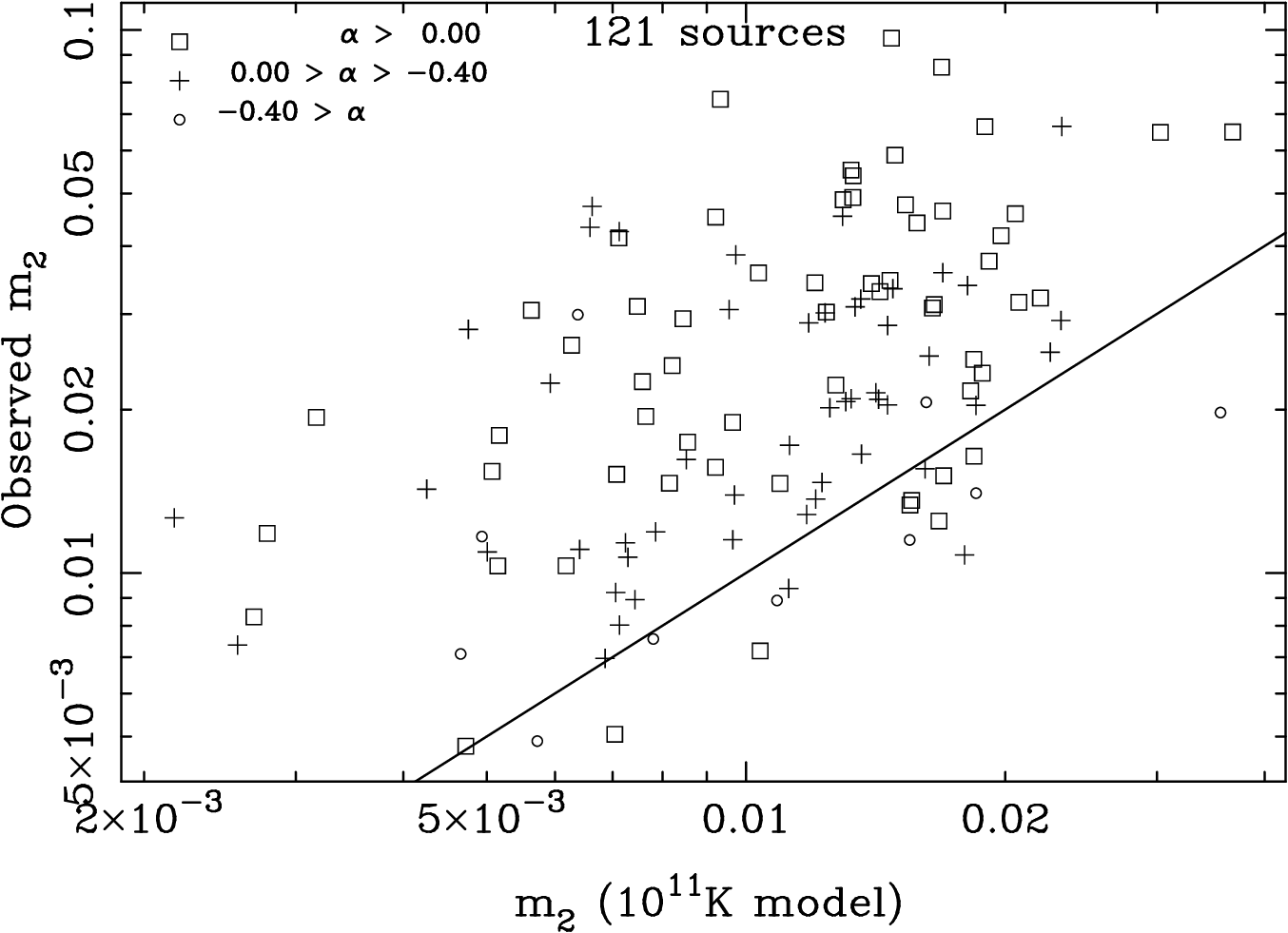} &
\includegraphics[width=5.5cm,angle=-90]{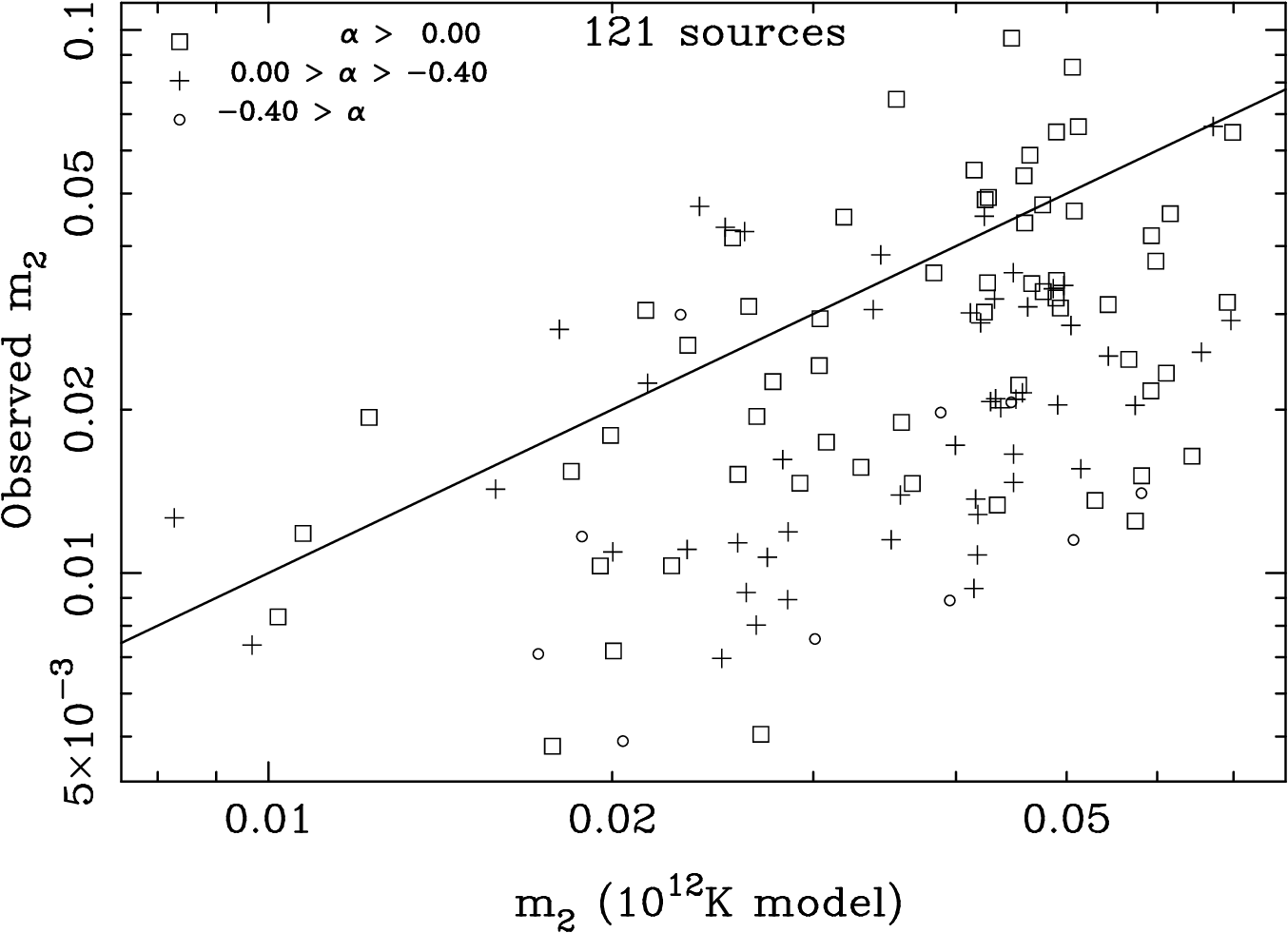}
\end{tabular}
\figcaption[]{Observed ISS index at~2~GHz plotted against
the model predictions. In the two models shown half 
of each source flux density is assumed to be a 
Gaussian ``core'' whose peak brightness is $10^{11}$~K 
and $10^{12}$~K  on the left and right, respectively.
The points are flagged by spectral index as indicated.
\label{f:m.mtheor} }
\end{figure*}

The right panel shows the subset of
sources with VLBI-observed structure index. The sources with structure
indices of 1 or 2 have mostly flatter spectra
than those with more extended structure. Also many of
the compact structure sources show scintillation indices above
2\%.  While a structure index of 1 or 2 does imply a small diameter,
we note that structure index is defined as a measure
the simplicity of a source's fine structure, that is partially 
resolved in VLBI; larger structure indices indicate
complex structures which can confuse the use of VLBI observations for 
geodetic purposes. 
Thus it is not a simple measure
of source angular diameter nor of the
flux density in a source that is unresolved in a VLBI observation.
We have also examined a similar plot flagged by
the VLBI model source diameters mentioned in section
\ref{sec:observe}, but there is not a clear trend of
$m_2$ with these diameters.

Figure~\ref{f:m.mtheor} compares the observed scintillation indices
directly with those predicted from the $10^{11}$~K and $10^{12}$~K models. 
In this and subsequent figures, except where noted, 
we include only those sources with 2~GHz ISS confidence levels 
of 1 and 2 and spectral index $\alpha>-0.6$.
The left panel shows that the $10^{11}$~K model correlates 
weakly with the observations, with observed
$m_2$ values lying mostly above the predictions.
Note that more sources with flat or inverted 
spectra tend to lie on the high side of
the prediction, as expected.  In the right panel
the observed $m_2$ values lie mostly below the predictions
for the $10^{12}$~K model.
We conclude that typical brightness temperatures
for the sources are in the range $10^{11}$~K--$10^{12}$~K.

The lack of a tight correlation is presumably due to 
departures from our simple source model. One cause of scatter
would be a distribution in the fraction of flux density in the 
compact core.  Consider the effect of a brighter core with a 
smaller diameter (following equation \ref{eq:FWHM}).
This would increase the scintillation index, but in order to 
match an observed $S_{\rm rms}$ and fixed parameters of the 
scattering medium, we would need a reduced 
flux density in the compact core. 
Quantitatively, an observed $m_2$ compatible with 50\%
flux density in a $3\times 10^{11}$~K 
component could also be modelled as $3\times 10^{12}$~K
component with about 5\% of the total flux density, 
providing that the source diameter remains large
enough to partially quench the scintillations.
For a source small enough to scintillate like a point source the 
lowest value for ${S_c}$ becomes equal to the observed 
$S_{\rm rms}$ which is also a few percent of the total.

\begin{figure}[htb]
\begin{tabular}{l}
\includegraphics[width=5.5cm,angle=-90]{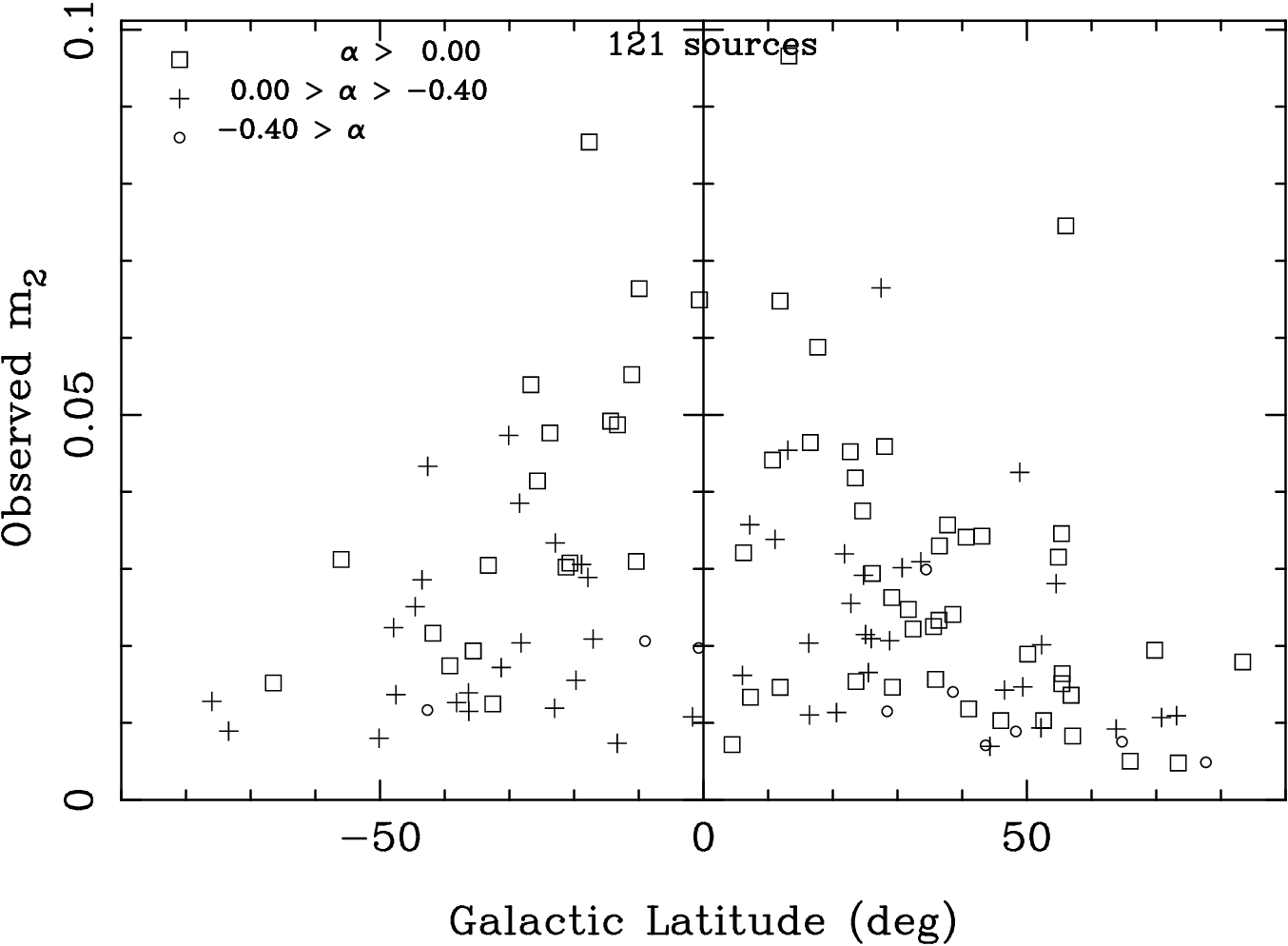} \\
\includegraphics[width=5.5cm,angle=-90]{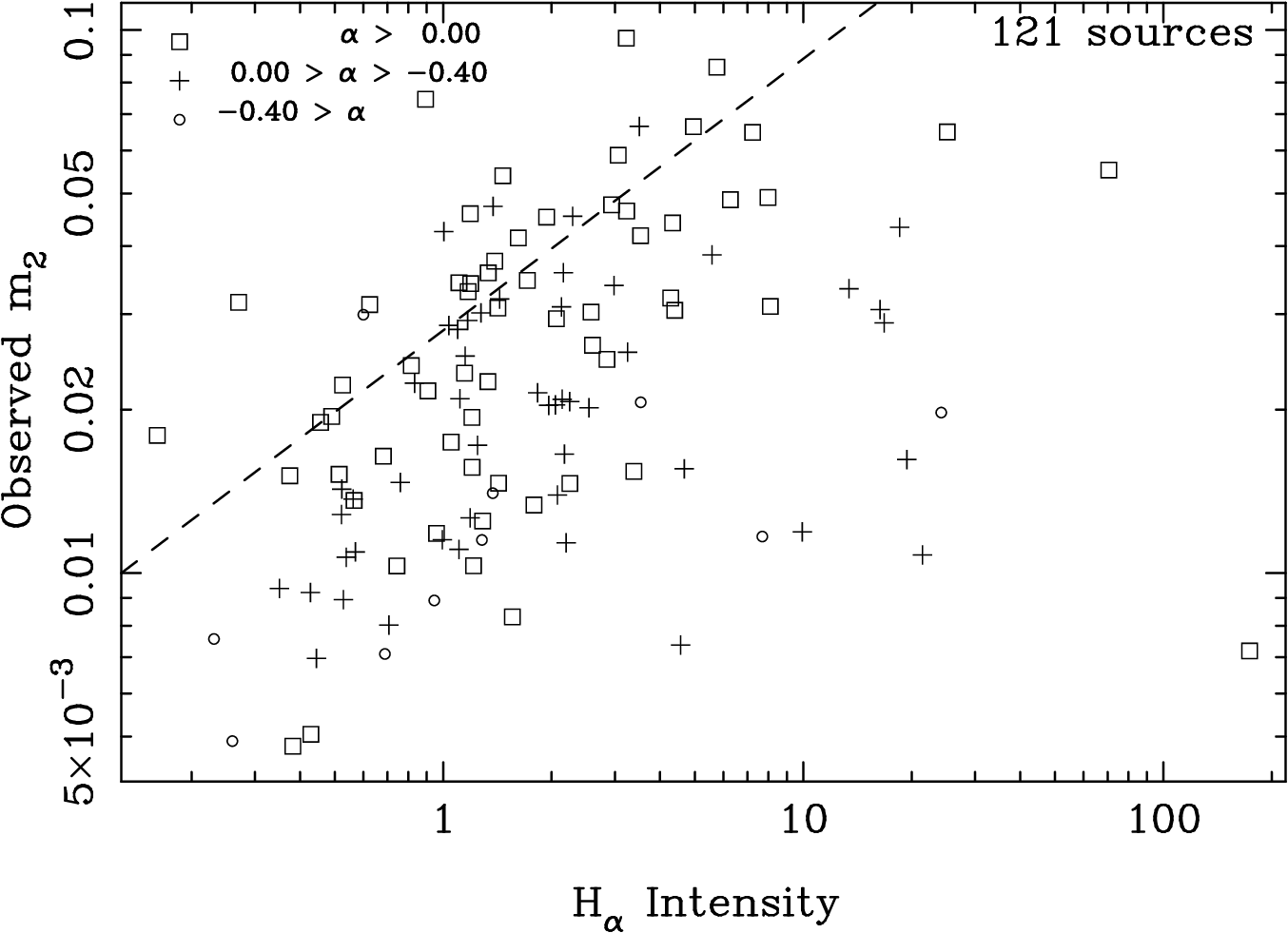}
\end{tabular}
\figcaption[]{Scintillation index at 2 GHz 
(for sources with ISS confidence levels 1 and~2)
plotted against  \textit{Left}: Galactic latitude and 
\textit{Right}: H$\alpha$ intensity ($I_{\alpha}$ in Rayleighs) 
as explained in the text.
The points are flagged by spectral index as indicated.
The dashed line corresponds to $m \propto \sqrt{I_{\alpha}}$.
\label{f:m_b} }
\end{figure}

The left panel of Figure~\ref{f:m_b} 
shows $m_2$ as a function of Galactic latitude,
flagged according to $\alpha$.  There is a generally similar 
dependence of $m_2$ on $b$ as in the model
of Figure~\ref{f:mt.theor_b}.  Again the influence of source diameter is 
shown by the trend for more flat spectrum sources to have higher
$m_2$ values.  We examined a similar plot for the 56 sources 
with credible values for $m_8$ but it showed only a slight 
dependence on latitude.  
The right panel shows $m_2$ plotted against $I_{\alpha}$,
where $I_{\alpha}$ is the intensity of the H$_{\alpha}$
emission as observed by the Wisconsin
Hydrogen Alpha Mapper in the one degree beam nearest 
that direction (Haffner et al., 2002).  $I_{\alpha}$ 
(integrated over -80 to 80 km s$^{-1}$) is 
measured in Rayleighs and can be interpreted as proportional 
to the emission measure in the warm ionized ISM (1 Rayleigh
corresponds to an emission measure of about 
2.25 cm$^{-6}$~pc at about 8000 K);
thus it is an independent measure of the column integrated
electron density squared.  There is a clear trend
of increasing $m_2$ with  $I_{\alpha}$, and the trend is 
strongest for the most compact sources 
(flat or inverted spectra shown as squares). 
The dashed line $m_2 \propto \sqrt{\rm SM} \propto \sqrt{I_{\alpha}}$ gives
the theoretical relation if emission measure is 
proportional scattering measure ${\rm SM}$. This 
applies if the electron density spectrum follows a power law
with an outer scale which determines the density variance,
which in turn is proportional to the square of total 
electron density (see CL05 for example). The dashed line is placed
to approximate the apparent upper limit line for $m_2$.

\begin{figure}[htb]
\includegraphics[width=5.5cm,angle=-90]{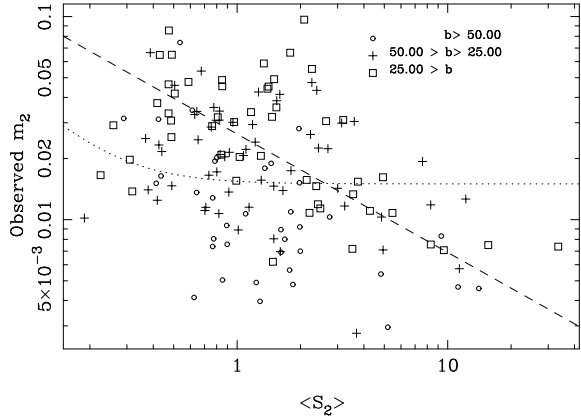}
\figcaption[]{Scintillation index observed at~2~GHz against mean 
source flux density.  The different symbols flag ranges of the
magnitude of the Galactic latitude.  The dotted line is
from equation (\ref{eq:sigma2}) and the dashed line is the rough
lower limit for the $10^{12}$K model from Figure \ref{f:mt.theor_b}.
\label{f:m_sbar} }
\end{figure}

\begin{figure}[htb]
\includegraphics[width=5.5cm,angle=-90]{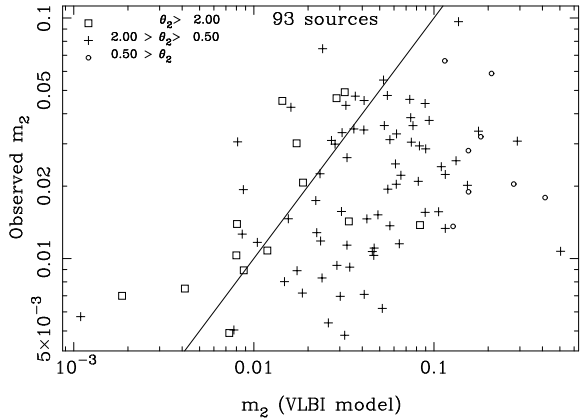} 
\figcaption[]{Observed $m_2$ against 
predicted from the ISS model using source data from S-band VLBI.
The points, flagged according to the source diameters, only show
agreement for the larger sources.
\label{f:m_2.vlbi} }
\end{figure}

Figure~\ref{f:m_sbar} shows the influence 
of $\ave{S}$ on $m_2$. There is a 
deficit of points in the upper right and lower left
of Figure~\ref{f:m_sbar}, but we do not see the 
strong inverse relation of $m_2$ and $\ave{S}$ visible in 
the model of Figure \ref{f:mt.theor_b}.
This suggests only a weak influence of $\ave{S}$ on source diameter.
Is this due to errors or due to inadequacies of the model?
The dotted line is the single point error (normalized by $\ave{S}$) 
from equation~(\ref{eq:sigma2}). Though in an ideal 
observation one could detect variations well below
this line, as noted in \S\ref{sec:method} we consider a lower
limit for a credible $m$~value to be $\sim$1\%.
Since the observations extend up to nearly 10\%, 
most of the points are not much influenced by errors. 
The spread must be caused by a combination of the range of scattering 
conditions, as indicated approximately by the Galactic latitude
and the range of actual source brightness. 
The dashed line is $m_{\rm 2} \propto \ave{S}^{-7/12}$
from figure \ref{f:mt.theor_b},
which approximates the $10^{12}$K model.
As noted earlier it is likely that there is
a wide range in the fraction of the total flux density 
in the compact (scintillating) components, which
contributes to the scatter about this inverse relation 
of $m_2$ and $\ave{S}$.

In Table~\ref{tab:results} we list the 
flux density and FWHM diameter of the most compact component 
derived from model fits 2.4 GHz VLBI observations.  This
gives compact structural information on
93 of our sources (with diameters from 0.3 mas to 8 mas).
We tried these values in place of the brightness-limited
models to calculate $m_2$ and $\tau_2$.
Figure \ref{f:m_2.vlbi} shows the observed $m_2$ versus
this prediction, flagged by the FWHM $\theta_{2}$. 
Most of the predicted $m_2$ values are larger
than observed;  but for the diameters larger than about 2 mas
the observations approximately follow with the predictions.
A scatter plot of the 93 VLBI diameters against flux density
shows no inverse relation expected if the structures were
brightness-limited. It seems that extracting the single
most compact component in model fits to VLBI data is
not the best way to extract compact structural information
for comparison with scintillation data, since sources typically show
structure on a range of scales. We conclude that the 2.4 GHz VLBI
observations are of limited utility, except for the more resolved
sources.

\subsection{A Global Comparison for scintillation timescale}
\label{sec:global.tau}

\begin{figure}[htb]
\begin{tabular}{l}
\includegraphics[width=5cm,angle=-90]{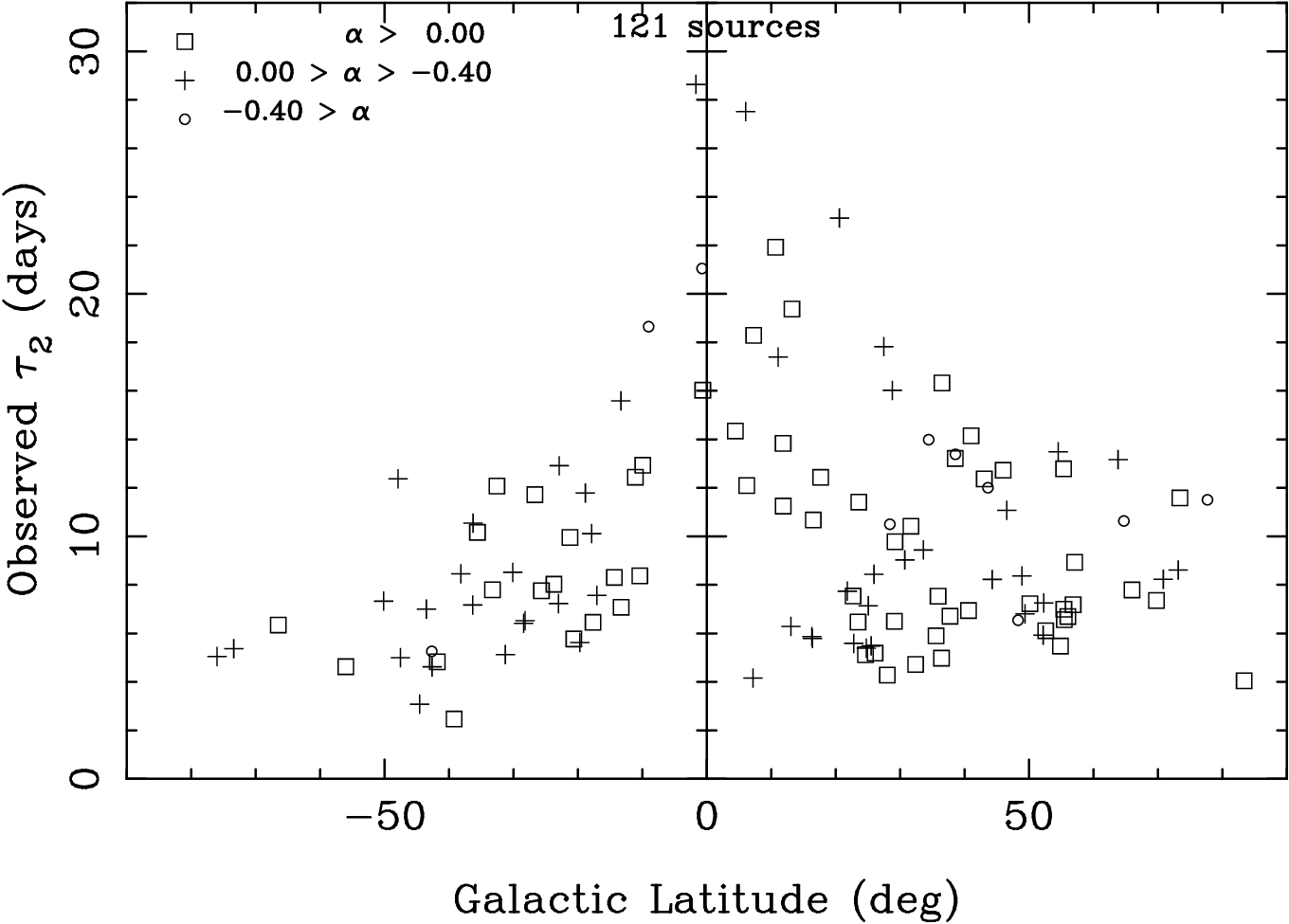} \\
\includegraphics[width=5cm,angle=-90]{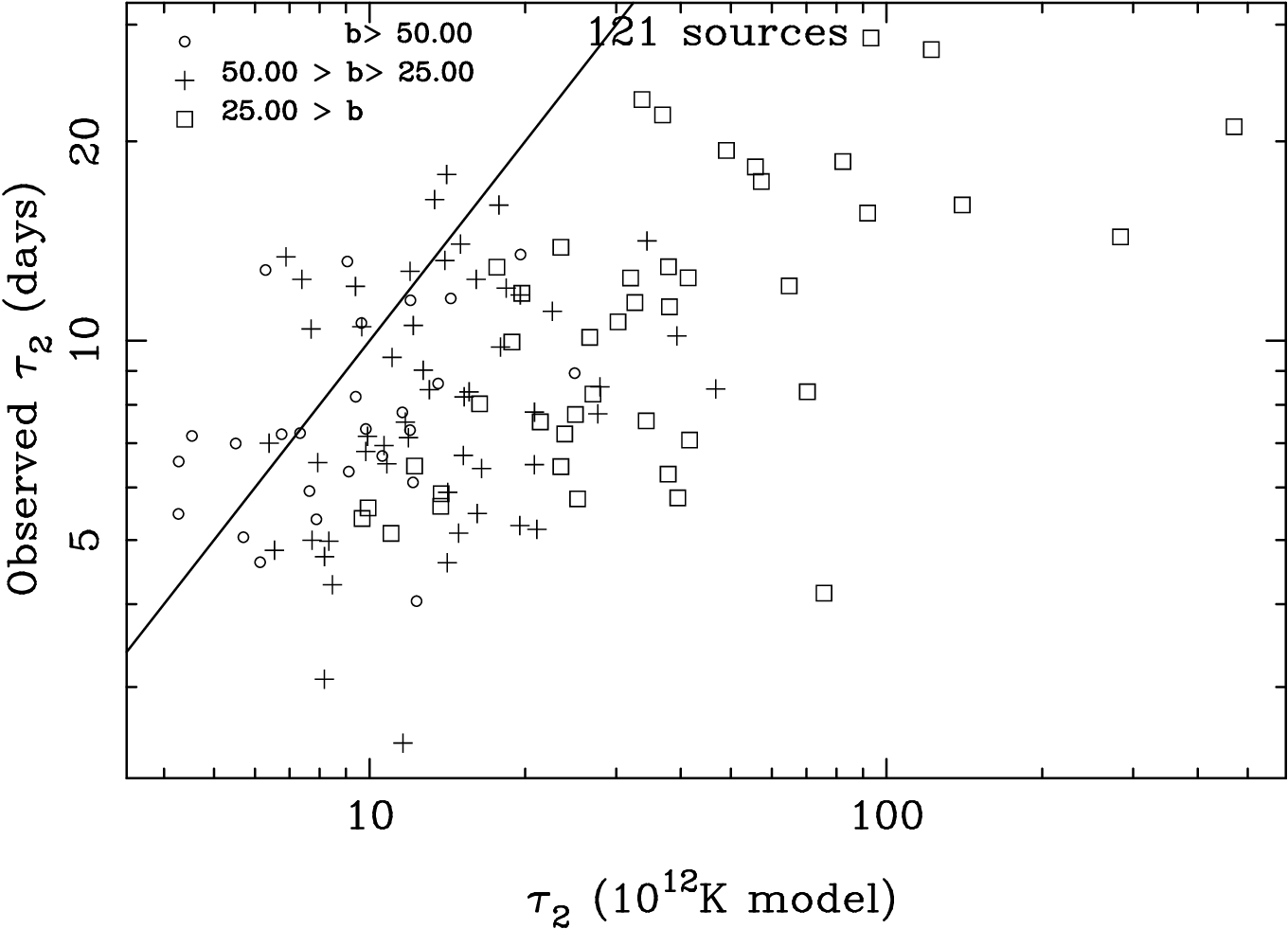}
\end{tabular}
\figcaption[]{\textit{Left}: 2~GHz ISS timescale versus Galactic latitude,
flagged for the indicated ranges $\alpha$. At a given latitude
we expect shorter timescales from the more compact sources
which typically have higher values of $\alpha$. 
\textit{Right}: 2~GHz ISS timescale versus model $\tau_2$ 
for the $10^{12}$~K source model, flagged for the Galactic
latitude as indicated.
\label{f:t_b} }
\end{figure}

We now consider the 2~GHz ISS timescale~$\tau_2$.
Figure~\ref{f:t_b} shows observed 
$\tau_2$ as a function of~$b$ in the left panel. Whereas 
there is clear increase in $\tau_2$ at low latitudes, 
the increase is much less than for the models,
as shown in the right panel which plots the observed 
against the predicted timescale.
At latitudes below about 10 degrees the model timescales rise 
to more than 500 d, due to the very
long scattering length through the inner Galaxy.
The timescale $\tau_{\rm iss}$ is given 
by the spatial scale $s_{\rm iss}$ divided
by the Earth's velocity with respect to
the scattering plasma.  In our model we took the
velocity to be the yearly average of the Earth's
transverse speed with respect to the LSR.  Whereas 
this may be reasonable for scattering paths out to
a few hundred parsecs, it will systematically
underestimate the velocity at greater distances,
due to the absence of differential Galactic rotation
and the peculiar motions of the ionized ISM
in the model.  This presumably 
explains at least part of the discrepancy.

Overall, we conclude that the general behaviour of the scintillation index
$m_2$ at 2~GHz is quite well described by the CL05 model of the interstellar
electrons with the flat spectrum sources having
a substantial fraction of their flux density in regions
of $3 \times 10^{11}$K peak brightness.  Ten times higher brightness 
is possible only from sources with less than about 5\% of 
their total flux density in the most compact core.
Our analysis shows only a weak inverse relation of source 
size to flux density. Thus the sources must be exhibiting a range
of peak brightness temperatures around $3 \times 10^{11}$K, as could
be explained by a spread in their Doppler factors.

\subsection{Comparison of Fast and Slow Variations}
\label{sec:vary}

\begin{figure}[htb]
\includegraphics[width=5.5cm,angle=-90]{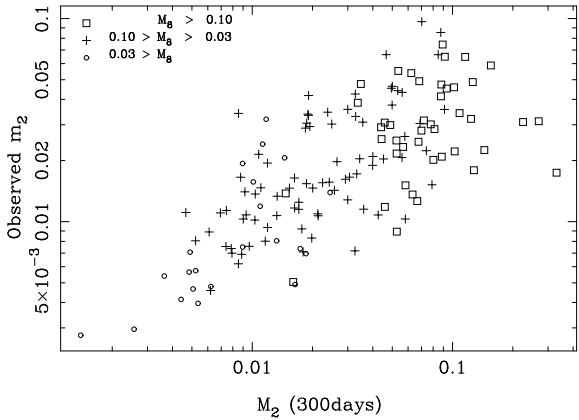} 
\figcaption[]{2~GHz comparison of short-term 
and 300~day modulation indices observed at 2 GHz, flagged by
300~day modulation indices at 8 GHz. 
\label{f:m_m300s} }
\end{figure}

As already noted there are both intrinsic and ISS variations in 
many of the GBI time series.  Indeed the presence of intrinsic (slower) 
variations was part of the selection criteria for the sample.
Since  slow intrinsic variations and fast ISS variations 
are both indicators of compact source structure, 
we examine the relationship between them.

As described in \S\ref{sec:method} we characterize
the low-pass filtered time series by $M_2$ and $M_8$,
which are the rms variations on a timescale of 300~days
(normalized by the mean flux density), as
given in Table \ref{tab:results}.
We note that these quantities respond to variations
on scales of 300-days or longer, including linear trends across
the entire data span.
Figure \ref{f:m_m300s} shows the 2~GHz
ISS index $m_2$ against $M_2$, flagged by $M_8$.  
In this plot we included all
the 146 sources, to avoid a bias against low $m_2$ values
for cases where the 2 GHz ISS was not reliably estimated
(ISS confidence level 0). The plot shows a strong correlation of
the incidence of ISS and intrinsic variations at 2~GHz.
This is consistent with both ISS and intrinsic variations
being associated with small diameters. 
The diagram also shows intrinsic variations to be 
strongly correlated between 2 and 8~GHz as expected.

We also note that Hughes, Aller, \& Aller~(1992)
found time-scales and slopes for the cm-wavelength
variation from a set of sources monitored with
the 26~m antenna at the University of Michigan.
27 of their sources are in the GBI source list.
Their time-scales range from 0.3 to over 3 years, and their 
structure functions on these time-scales were clustered
about 1 year, which they ascribe to intrinsic
variation.

\begin{figure*}
\begin{tabular}{lr}
\includegraphics[width=6cm]{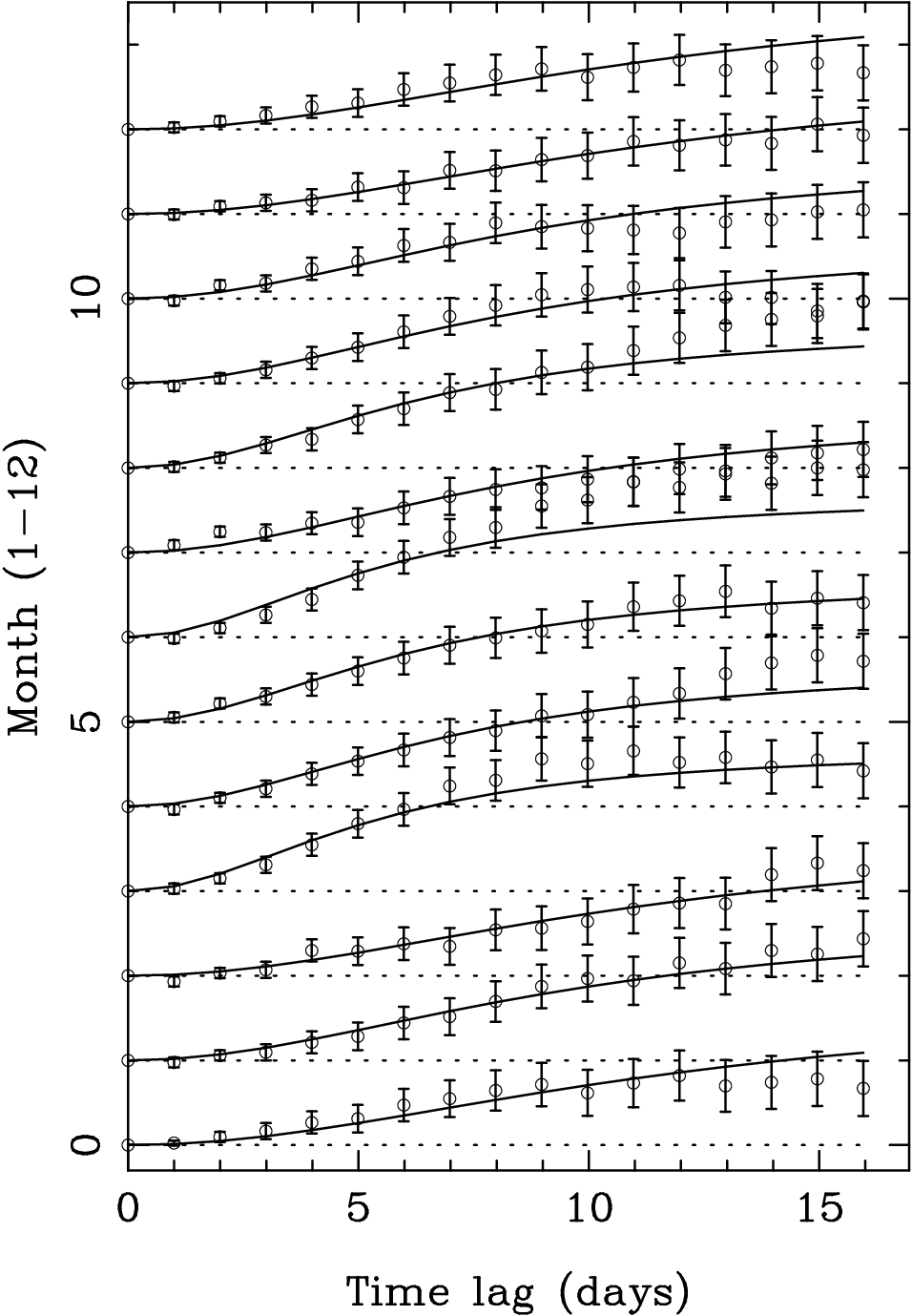}  &
\includegraphics[width=6cm]{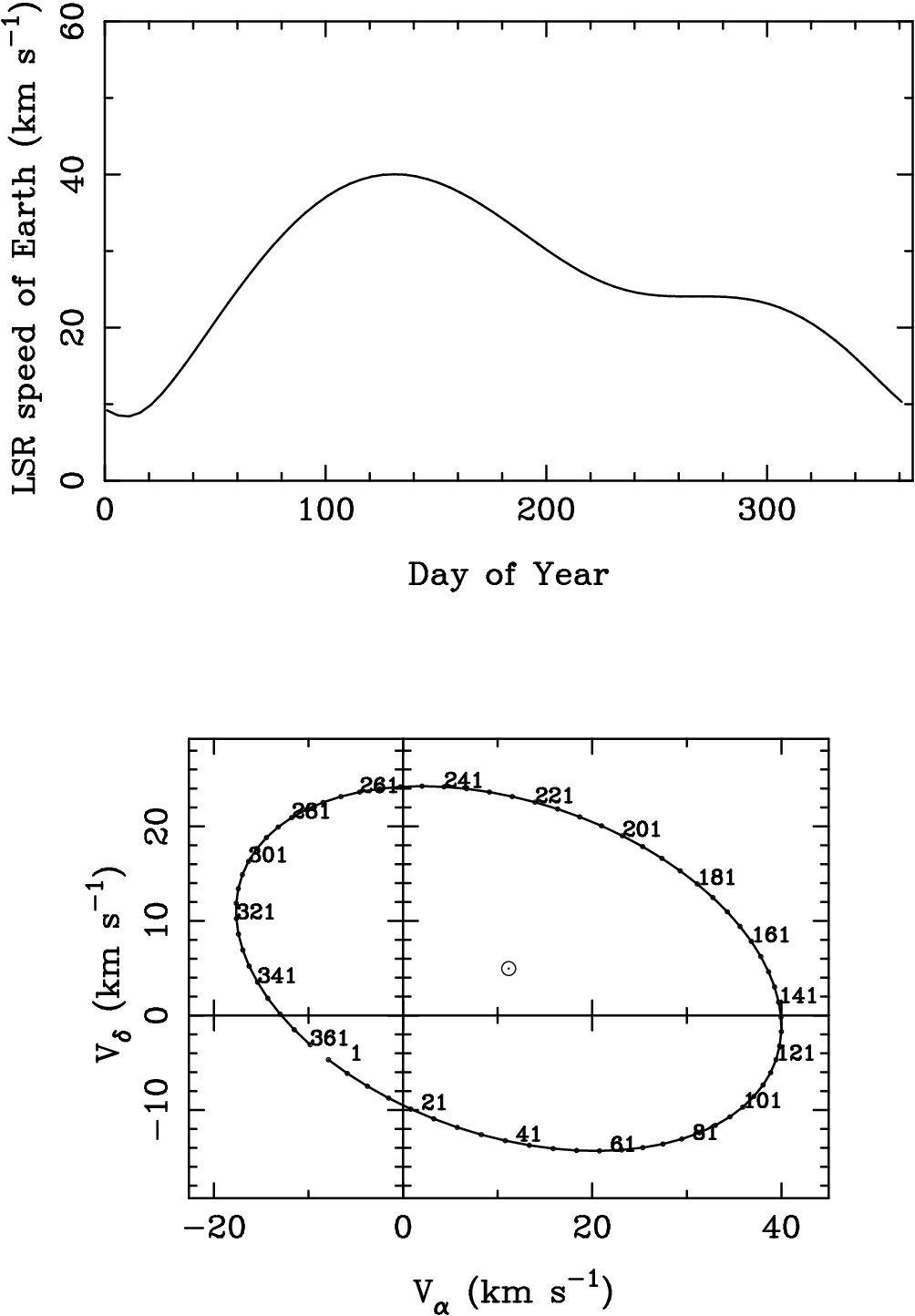}
\end{tabular}
\figcaption[]{\textit{Left}:
Monthly 2~GHz structure function estimates for 1514$+$197.
December is repeated as month 0 and 12.
\textit{Right}: Earth velocity relative to LSR 
perpendicular to the
line of sight toward 1514$+$197 ($V_{\rm LSR}$)  
against day of year.
\label{f:strfn12.1514} }
\end{figure*}

\subsection{Annual Modulation in ISS time-scale}\label{sec:annual}

It has been established that there are annual
modulations of the characteristic timescale in several IDV sources 
(J1819$+$385, 0917$+$624, 1257$-$326 1519$-$273: Dennett-Thorpe and de Bruyn, 2003;
Rickett et al., 2001; Jauncey and Macquart, 2001; Bignall et al., 2003;
Jauncey et al., 2003). 

The modulations are due to the annual cycle of the Earth's velocity
relative to the scattering medium. Accordingly, we have looked for
similar annual modulations in the ISS signal at~2~GHz. 
The right panels of Figure \ref{f:strfn12.1514} show the predicted
annual velocity variation for blazar 1514+197.

Our analysis consisted of computing 
structure functions out to lags of 30 days
averaged for each calendar month over all years available.  
These 12 estimates and their errors are shown
by the points in the left panel of 
Figure \ref{f:strfn12.1514} for 1514$+$197.  
(December is repeated as month 0 and 12). 
From plots such as this we estimated 12 timescales $t_o$,
by fitting smooth models to each monthly structure funtion.
We first fitted a form to the structure function
of the entire time series and selected the best
fit as defining the functional form to be used for each month.
We tried various forms including an exponential and functions
of the form:  
\be
D_m(\tau) = D_S \frac{t^p}{t^p + t_o^p} .
\ee
We optimized the parameters $D_s$, $t_o$ and $p$ (in the range 
1.4 to 2.0) and also the exponential (used in place
of p=1) and selected the best fit.  
With $D_s$ and $p$ then fixed, we found the best-fitting value
of the time scale $t_o$ for each of the 12 months for
that source.  When avaliable we did independent analyses for
the 1979--87 and 1988--96 data sets.

\begin{figure}[htb]
\includegraphics[width=8cm]{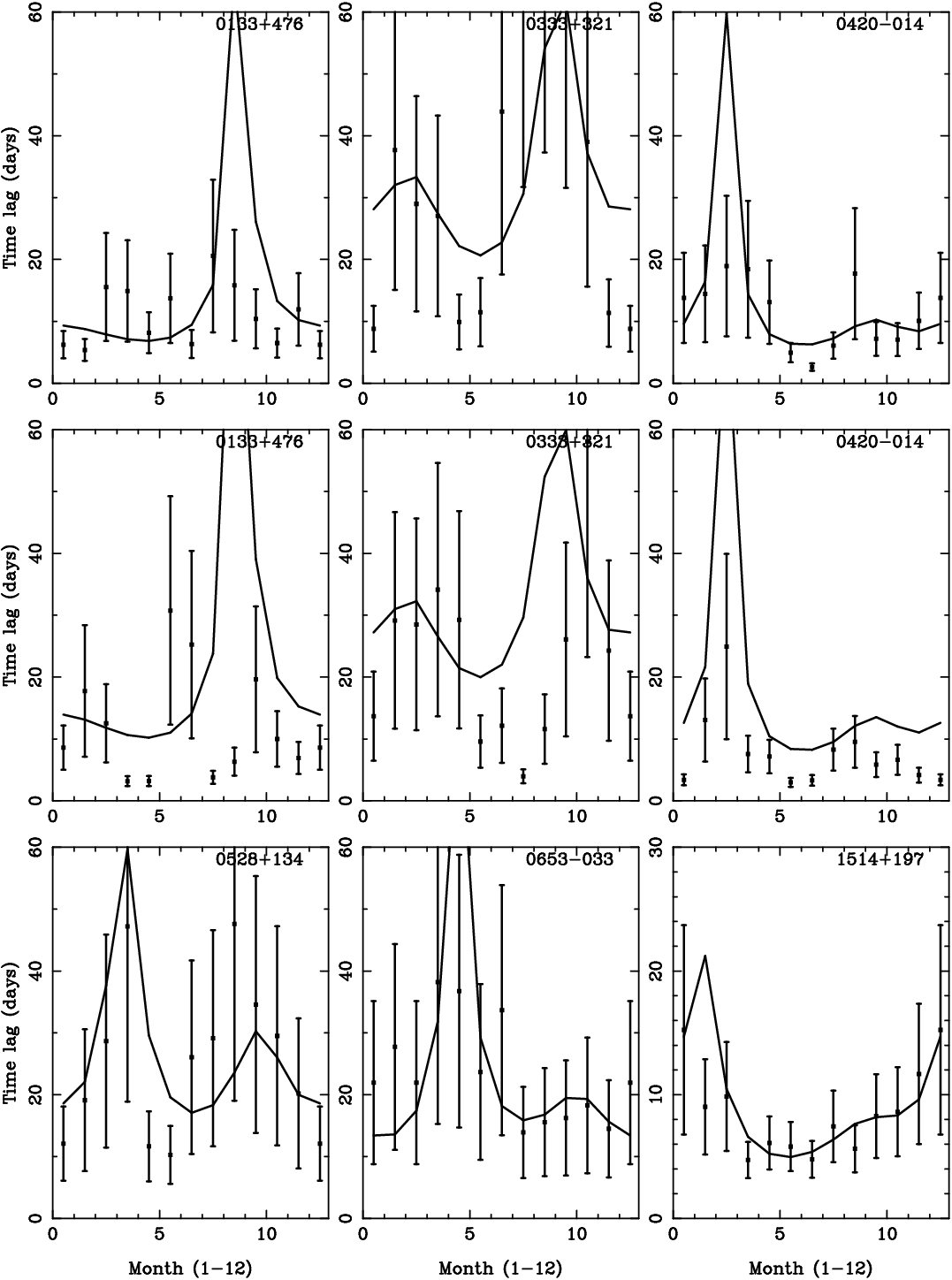} 
\figcaption[]{Monthly 2~GHz timescale estimates for six
sources: top panels from 1979--87; center and bottom
panels from 1988--96. Sources 0133$+$478, 0333$+$321 and 0420$-$014
were observed in both periods. The points show observations;
the line shows the LSR prediction.
\label{f:annmod} }
\end{figure}

Figure \ref{f:annmod} shows some results as nine plots
for six sources. The best example is 1514$+$197 in the 
lower right panel. The points are $t_o$ and their errors; the lines are
predictions ($\propto 1/V_{\rm LSR}$), assuming that the scattering medium 
is stationary in the LSR, where the constant of proportionality
was set to give one half of the maximum observed $t_o$ for the annual
average of $V_{\rm LSR}$.  For 1514$+$197 there is good 
agreement between the LSR-model and the observations.
See also the right panel of Figure \ref{f:strfn12.1514}. 
However, for the great majority of the other sources there is not 
a convincing trend of $t_o$ over the year (relative to the errors);
even though, with many years of data, we have as many as 
7--8 independent month-long samples for over 100 sources.

There are several factors affecting the analysis.
First, the reduced signal-to-noise in one twelfth of the data;
second, when $\tau_2 \simgreat 20$ days there is only one
independent ``scintle'' in each month; third, for low latitude sources
the long scattering path makes for large differential motion 
(systematic and random) through the Galaxy,
which will tend to dominate the $\sim$30 km s$^{-1}$
variations of the Earth; fourth, for several sources there are large 
changes in the flux density over one-month timescales
which are, presumably, intrinsic and may correspond to
changing fine structure in the source on similar timescales.
All of these factors will confuse the systematic
annual variation.  Further the phase (and shape) of the
annual curve depend on both the ecliptic coordinates of the source
and on the LSR velocity of the scattering plasma.
Though the IDV in sources 0917$+$624 and 1257$-$326 were well
described by the LSR-model, the IDV of J1819$+$385 
requires a velocity of about 20 km s$^{-1}$ relative to the LSR.
Thus our predicted curves are only a first-order model for the
ISS timescale.

Panels 1-6 of the Figure \ref{f:annmod} show three sources
(0133$+$478, 0333$+$321 and 0420$-$014)
observed in both 1979--87 and 1988--96. They were selected
as the sources observed in both periods 
with $m_2 \simgreat 0.03$ and in which there was some appearance
of an annual modulation in agreement with the predictions.
Comparing the upper and central panels the data points 
show some relation to the predictions in both epochs for each source.
However, individual months have
significant differences between the epochs. The lower panels show the
sources 1528$+$134, 1653$-$033 and 1514$+$197, which are the best
examples of annual cycles seen in the 1988--96 data set.

\subsection{The ISM model}
\label{sec:ISM-model}

\begin{figure}[th]
\includegraphics[width=6cm]{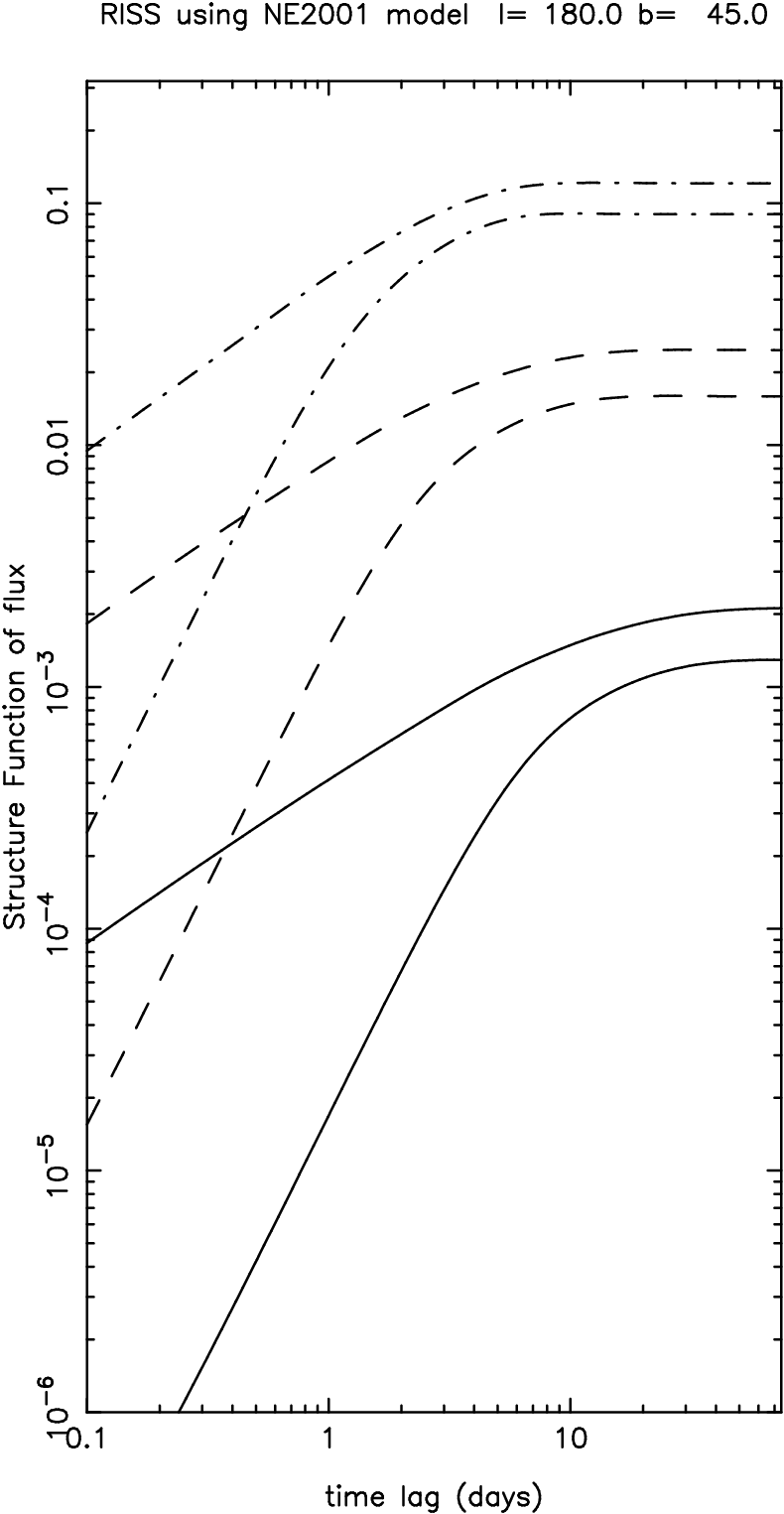} 
\figcaption[]{Theoretical ISS structure functions for sources with 0.5 Jansky
in a Gaussian core of brightness $10^{11}$K (solid), $10^{12}$K (dashed)
and  $10^{13}$K (dash-dot).  Calculations are for 180\de Galactic longitude and
45\de latitude with the CL05 electron model. The lower steeper curves include a local bubble with much reduced electron
so that the electron distribution extends to the solar vicinity.
\label{f:local_bubble} }
\end{figure}

In the foregoing we have shown that the rapid 2~GHz flux density 
variations in the GBI data are generally consistent with predictions
from the CL05 model for the electron distribution in our Galaxy,
assuming models for the most compact components in the 146 
extragalactic radio sources that do not require extreme Doppler boosting
factors and are in general consistent with the maximum brightness 
obtained from VLBI.  We have also compared the observations with
predictions from the TC93 electron model.  The predicted scintillation 
levels are somewhat higher from the TC93 model, such that the typical
implied source brightness is $10^{11}$K, i.e.\ a factor three below
that implied using the CL05 model. This difference highlights 
the importance of refining our knowledge of the electron distribution
in the estimation of source structure from ISS.

As mentioned in the introduction variation on timescales 
as short as one hour are seen from a few extragalactic sources at cm wavelengths. 
Such rapid variations can only be explained by local scattering ``clouds'' located
10-30 pc from the Earth.  This is in marked contrast to our results
reported here in which the CL05 model can successfully explain 
the 2~GHz variations from 121 comapct radio sources.  The CL05 model
involves scattering at typical distances of 500-1000~pc and
has a deficit of electrons in a ``local bubble''; thus it
specifically excludes scattering in the local ISM.  The importance of this local
bubble on the predicted ISS is illustrated in Figure \ref{f:local_bubble}.
Here the lower curves have reduced scattering plasma in the local bubble
(i.e.\ within about 100 pc of the Sun).  Related plots
were made by Blandford, Narayan and Romani (1986) and by Coles (1988). 
The figure shows that observations
of the logarithmic slope of the structure can be used to test for
scintillation caused in the local ISM; slopes as steep as two
are expected when the local bubble eliminates any local scattering,
and slopes of 0.67 are expected if the scattering extends to near the Sun.  
In L01 the slopes estimated from the 2~GHz GBI data were found to lie in the range
0 to 1.2, however, these estimates were not corrected for
noise and so cannot be used for this purpose.  We do not think that there
is a sufficient range of timescales in the data to reliably determine these
slopes.  The plots of structure function $S_{I}(T)$
are useful in predicting the apparent scintillation index {$m$} in flux density 
over an observing span $T$, as $m \sim \sqrt{0.5 S_I(T)}$.  
The differences in Figure \ref{f:local_bubble} largely explain the 
higher predicted scintillation indices from the TC93 model, which does 
not model the local bubble. 

An important result bearing on the existence of local clouds 
comes from the MASIV survey reported by Lovell et al. (2003).
They observed 710 flat spectrum sources at 4.9 GHz for 72 hours
and found 85 with significant variation over that time. However,
they only detected one source with variations as fast as 4 hours,
which was the well-known IDV quasar J1819+385. The sources surveyed
covered most of the Northern sky and so suggest that the covering fraction of
local scattering clouds is small (1/85).  The GBI observations were not
sensitive to ISS on times shorter than one day, but the rarity of time-scales
in the range 2-4~d also suggests that the local ISM is indeed depleted
of scattering plasma and so supports the local bubble concept.

Another aspect of the three very rapid variables is that detailed modelling
of their ISS shows that their local clouds cause anisotropic scattering,
i.e.\ elongated plasma density structures presumably oriented parallel
to the local magnetic field. In contrast we assume the scattering to be
circularly symmetric in our modelling of the GBI data.  Essentially this 
assumes that the length scale for randomization of the projected magnetic field
is substantilly smaller than the 500-1000~pc 
distances through the ISM. However, there is insufficient
information in the GBI data to test this assumption. Anisotropy would cause
the structure functions (as in Figure \ref{f:local_bubble})
to have different widths and shapes in orthogonal directions.
These considerations should be included in analysis of any
rapidly sampled flux densities in future observations (such as
that proposed at the PisGah Observatory).

\section{Comments on Individual Sources}
\label{sec:individual}

We examined plots as in Figures \ref{f:flux1514} and
\ref{f:acf1514} for all of the 44 sources observed in
1979--87 and the 146 sources observed in 1988--96. 
Whereas the parameters derived from the correlation analysis are
listed in Table \ref{tab:results} and whereas most conform
to our hypothesis of rapid ISS at 2~GHz and slower intrinsic
variation at 8~GHz, we now discuss some sources  
which had unusual time series or were otherwise notable.
The interested reader is encouraged to peruse the 146 
time series plots in L01; where there is also a list 
of identifications and alternate names for the sources.

Among the sources discussed individually in this section
7 out of 12 are BL Lac objects, compared with 20 BL Lacs
among the 146 sources monitored. This may suggest that
unusual ISS or unusual intrinsic variations are more common
in the BL Lacs than in the quasars, but such behaviour
was part of the selection criterion for the observations.

\subsubsection*{0016$+$731}

This quasar shows obvious variations ($m_2 = 5$\%) 
on a 1 month timescale at 2~GHz with no slow variations at 8~GHz.  
The 2~GHz variations appear to be relatively slow ISS; 
at a latitude of 11\de\ it is near the upper envelope of the 
$\tau_2$ points in Figure \ref{f:t_b}. It has a relatively
high effective scattering distance of 2.55 kpc and so is probably exhibiting
refractive ISS.  At 8~GHz there may also be ISS variations 
on a timescale of 11 d.

\subsubsection*{0133$+$476}

The quasar 0133$+$476 is an excellent example of rapid ISS at 2~GHz and 
year-long intrinsic variations at 8~GHz. It was sampled daily in 1979-86 
and every other day in 1988--96. In both 2~GHz data sets there are 
stochastic variations by 10\% on timescales of 10 days; 
these have much lower amplitudes at 8~GHz but are revealed
by cross correlation. The 10-day timescale of its 2~GHz ISS
is short considering its Galactic latitude of -14\de.
The 8~GHz data show 25\% variations (presumably intrinsic) 
on timescales of a year, having a much lower amplitude at 2~GHz.

\subsubsection*{0235$+$164}

The BL Lac object 0235$+$164 is a much-studied variable at
many radio wavelengths.  Both the 2 and 8 GHz time series
are dominated by deep ($M_2 \sim .4 ; \; M_8 \sim 0.6$) 
relatively slow variations, which 
we interpret as intrinsic. The major outbursts over year-long
timescales are often accompanied by lower level events on times
of a few months. The 2~GHz variations are 
delayed by 1-3 months, with some evidence 
that the delay increased between 1982 and 1994.  
At 2~GHz there are also low level rapid variations particularly near the
peaks in the outbursts (e.g.\ in 1992--93).  We presume that these 
latter were ISS due to the emergence of new components
(like those in VLBI images) compact enough to 
scintillate, which then expanded and quenched the ISS.

\begin{figure*}
\includegraphics[width=7.5cm,angle=-90.0]{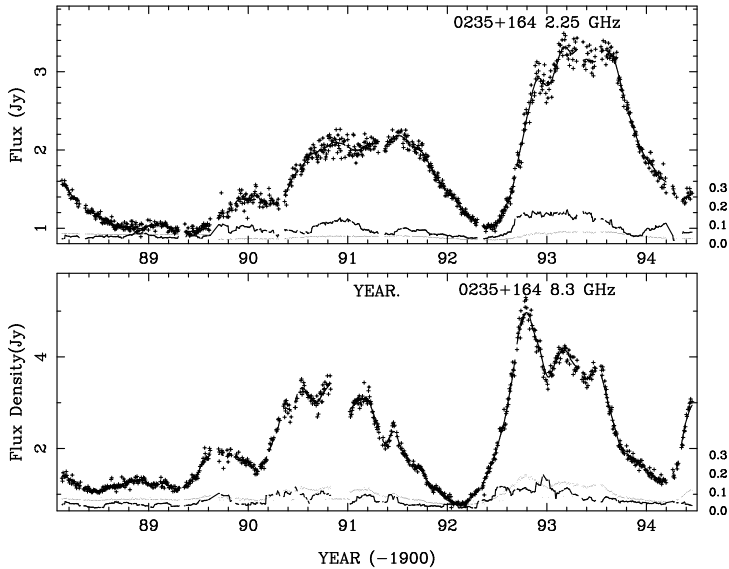} 
\figcaption[]{Flux density vs time for 
0235$+$164. The middle curve is a 50--day running estimate of
$\sigma_{\rm iss}$ described in the text and 
the lower curve is the rms error in flux density;
both curves scaled times 3 as indicated by scale on the right.
\label{f:flux0235} }
\end{figure*}

In Figure \ref{f:flux0235} we plot the flux densities
overplotted with a 50-day high pass running mean. The middle curve 
is a running estimate of $\sigma_{\rm iss}$, which is the net rms 
deviation in flux density from a quadratic curve fitted 
to the 50-day interval centered on each point.
At 2~GHz $\sigma_{\rm iss}$ is well above the lower curve which shows the 
rms due to noise and estimation error,
which has not been subtracted from $\sigma_{\rm iss}$. 
Often the ISS peaks earlier in a burst than does the 
total flux density, as expected for the ISS
of an emerging compact emission feature. A clear example
is the major burst during 1992--1993, where $\sigma_{\rm iss}$
increases abruptly in 1992.7 just as the 2~GHz flux density 
starts to rise. 
A simple model for the ISS of such a single transient component 
of diameter $\theta(t)$ and flux density $S_b(t)$ is
\be 
\sigma_{\rm iss} \sim S_b(t) m_o \theta_o/\sqrt{\theta(t)^2 + \theta_o^2}  ,
\ee
where $\theta_o$ is the critical diameter for suppression of the ISS
and $m_o$ is the scintillation index of a point source ($\sim 1.0$
at 2~GHz).  The physics of the expanding feature will determine
the relation of $\theta(t)$ to $S_b(t)$; while it is optically thick 
one may expect increasing $S_b(t) \propto \theta(t)^2$ 
and then a decay in $S_b(t)$ as it becomes optically thin. 

The 1992.7 event suggests a very rapid rise 
in $S_b(t)$ while its diameter  $\simless \theta_o$.
We shortened the time for the running estimate of 
$\sigma_{\rm iss}$ to 30-d and made a scatter plot of 
$\sigma_{\rm iss}$ against $S(t)$ during the rising part
of the burst. This revealed that $\sigma_{\rm iss}$ 
rose from 0.06 to 0.12 Jy while $S(t)$ increased from 
1.7 to 1.77 Jy. In other words
$\sigma_{\rm iss}$ increased by 0.06 Jy while the flux density
increased by 0.07 Jy (both figures have errors of about 25\%).
Thus 0.07 Jy flux density emerged with a scintillation index of 
$0.06/0.07 \sim 0.9$ implying a diameter $\simless \theta_o$.
which we estimate as about 10 $\mu$as using the $L_{eff}$ tabulated
in Table \ref{tab:results}. An emission component of
0.07 Jy with a diameter of 10 $\mu$as at 2.25 Ghz
implies a remarkably high brightness temperature 
of about $1.7 \times 10^{14}$K 
during the few days that $\sigma_{\rm iss}$ showed the rapid rise.
Subsequently $\sigma_{\rm iss}$ remained constant while the 
total flux density nearly doubled from 1.77 to 3.4 Jy.
This suggests a relationship between the effective diameter
of the new component and its contribution $S_b(t)$ to
the total flux density, but further theoretical work is needed to 
make a quantitative model of the expansion. 

It is notable that Kraus et al.\ (1999) were monitoring the
source at the VLA and observed a peak in the flux density
of 0235$+$164 at 1.5, 4.9 and 8.4 GHz on modified Julian
days 8904-8905, coincident with the 
abrupt increase in $\sigma_{\rm iss}$ at 2~GHz.  They considered
both intrinsic and extrinsic explanations, and our analysis suggests 
the emergence of a very bright component at that time, which contributed
to the rapid change in the light curve at 1.5 to 8.4~GHz.
However, the subsequent rapid variations of the flux density 
are clearly strongly influenced by ISS.  This event highlights
the importance of both phenomena in any detailed 
interpretation of such outbursts.

Viewed over the full span 1979-1996, the 2~GHz flux density for this 
source shows 30-50\% variations on times of several months combined with
only a low average level of ISS ($\sim 2$\%) variation on times of 3 days.
This is a lower level of ISS than seen for several sources
with very little intrinsic variaion (e.g.\ 0133$+$476 and 1514$+$197),
and it suggests that even though the source is highly variable
the average flux density coming from sub-mas structure is low
relative to that in sources which show a higher average value
for $m_2$.

\subsubsection*{0355$+$508}

This low latitude ($b= -1.7$\de) quasar (NRAO 150) was one of two
objects from the 1979--87 observations analyzed for ISS by Dennison 
et al.\ (1987).  At 2.7~GHz they found  a slow decay 
over 7 yr and 5\% variations on times of 1 yr, 
which they showed is consistent with
strong refractive ISS as expected at such a low latitude,
quenched by a 10:1 factor due to the angular diameter of 
the source which they estimated as 7 mas, which is comparable to the 5 mas
listed in Table \ref{tab:results}. Even at 
8.1 GHz the model predicts refractive ISS for this 
source. 

They made no precise definition of timescale
but 1 yr appears to be the ``quasi-period'' of the
variations about a slow decay, suggesting a timescale
in our definition of about 6 months. 
When analyzed through our 365-day high-pass filter
we find a time scale of 1-2 months, which illustrates
that the use of the high-pass filter to separate 
ISS and intrinsic variations is only possible when the two
timescales are widely different.  We detect 3\% variations
at 8 GHz on a timescale of 13 days, which is comaparable
to the 25 days predicted in their model.
 
\subsubsection*{0528$+$134}

As shown in figure \ref{f:flux0528}, this quasar 
was quiescent at 8~GHz over 1988--1992. It then started
a slow increase  about doubling its flux density by 1994. Then
over 1995.2--1996.0 it rose to 8 Jy, doubling its flux density again 
in less than 1 yr.   The ISS at 2~GHz is 5\% on a timescale of 12 d
persisting throughout the 8 years of monitoring.  We looked 
for the associated increase 2~GHz ISS amplitude
as the 8~GHz burst began, as seen in 0235$+$164. However,
though the flux density at 2~GHz did start to increase, 
the ISS amplitude did not, although it  
became somewhat faster. This shows that an increase in ISS is 
not always seen at the onset of an intrinsic burst.

\subsubsection*{0716$+$714}

0716$+$714 is a BL Lac object studied extensively 
by the Bonn group and exhibits IDV at cm wavelengths 
(e.g.\ see Quirrenbach et al., 2000).  Their structure 
function analysis shows significant variations on times 
2 days at the 1--2\% level. at 1.4 and 4.8 GHz.
In the GBI program it was mostly sampled at 2 day intervals 
and shows significant 2.2 GHz variation at about 4.8\% rms 
on a timescale of 5 days. 

Wagner et al.\ (1996) also report a relationship between 
its optical and radio variations, which they interpret 
as evidence for intrinsic IDV at cm wavelengths.  
If as they suggest the 4.8 GHz IDV were intrinsic 
with a Doppler factor $\sim 100$, the observable
angular diameter would be $\sim 3\mu$arcsec.
Scaling this to 2.2 GHz, we obtain $\sim 6 \mu$arcsec, which 
should scintillate as a point source, 
showing both refractive and diffractive ISS.
However, our 2.2~GHz observations suggest refractive ISS only
with possible smoothing by a source diameter of $\sim 0.5$ mas. 

\subsubsection*{0851$+$202}

Another BL Lac object 0851$+$202 (OJ287) is similar to 0235$+$164
in showing vigorous intrinsic variability on timescales of a 
few months, which is highly correlated between 2 and 8~GHz
and only 2.8\% rms in ISS at 2~GHz.  There are
peaks in the short-term rms in 2~GHz flux density at the beginning
of the intrinsic bursts, but the evidence is not as strong as
for 0235$+$164.

\subsubsection*{0954$+$658}

This BL Lac object shows 10-20\% variations over a year 
at 8 and 2 GHz, with the time delay from high to low
frequencies characteristic for intrinsic bursts.
It is well known since it produced the best and most 
definitive ESE in 1981 (Fiedler et al., 1987a). 
Our analysis also shows that it clearly exhibits typical
2~GHz ISS on a timescale of 4 days at 3.4\% scintillation index.


\subsubsection*{1413$+$135}

This BL Lac is remarkable for its strong intrinsic variations
(over  25\%) at 8~GHz and remarkably low 
level ISS ($\sim 0.6$\%) at 2~GHz, where it also has
only a low level of 300-day variation.  
Thus it appears as an outlier
near the center top of the right panel in
Figure \ref{f:m_m300s}.  This source has an inverted spectrum
($\alpha = 0.33$), making a possible explanation that
the compact core is heavily absorbed at 2~GHz
and contributes little to the 2~GHz flux density.

\subsubsection*{1514$+$197}

This BL Lac object shows the highest level
of ISS at 2~GHz (7.5\% rms on a tmescale of 7 d) 
of all of the extra-galactic sources
in the sample. At a latitude of 56\de\ it
is expected to be in weak ISS or near the transition
frequency. Thus its high level of ISS suggests a smaller diameter
than other sources.  2.3~GHz VLBI observations reported by
Fey and Charlot (2000) list its structure index as 1 --- the most compact
but its VLBI-modelled angular diameter in Table \ref{tab:results} is
1.8 mas which is not unusually small.
The relatively high level of ISS in this source revealed 
an annual modulation in its ISS timescale
agreeing in phase with that predicted for a scattering 
plasma at rest in the LSR.   We have computed the power spectrum
of its 2~GHz ISS and compared it with predictions for a Gaussian
source model, assuming a screen or the CL05 and TC93 models of the ionized
ISM.  These give agreement in the width at half power of
the ISS spectrum for a source diameter $\sim 0.25$ mas,
which corresponds to a brightness temperature of $\sim10^{12}$ K.  
As shown in Figure \ref{f:acf1514} and listed in Table
\ref{tab:results}, its ISS is also detected at 8 GHz with
an amplitude of 2\% on a timescale 11 d and is
52\% correlated with the 2 GHz ISS.  This is the best example 
of 8GHz ISS in our data.

\begin{figure*}
\includegraphics[width=8cm,angle=-90]{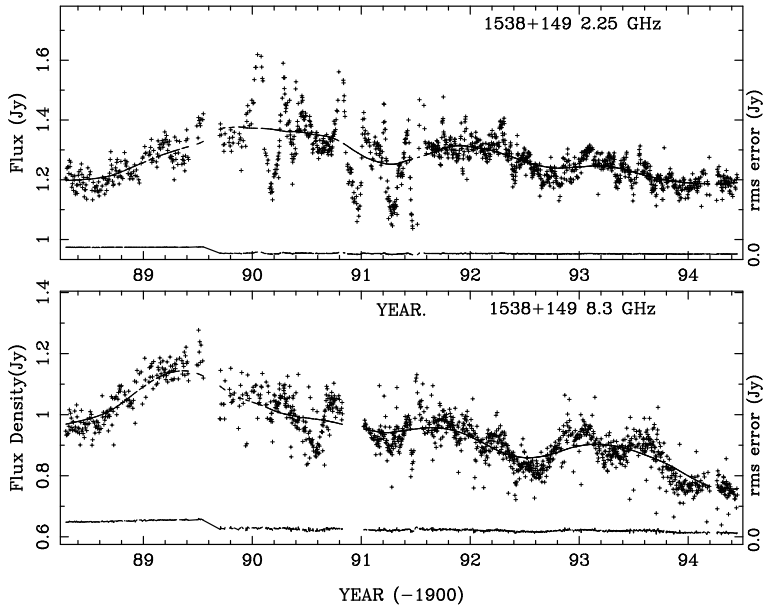}
\figcaption[]{Flux density at~2 and~8~GHz for 1538$+$149 in the same format
as Figure~\ref{f:flux1514}. Note the episode of enhanced variation at
2~GHz from 1989.9--1991.6, as discussed in \S\ref{sec:individual}.
\label{f:flux1538} }
\end{figure*}

\subsubsection*{1538$+$149}

The 2~GHz time series in Figure \ref{f:flux1538}
has an episode of strong variability
starting abruptly in 1989.9 and ending abruptly in 1991.6.
Before and after this episode there are typical
ISS variations at a few percent over 10 days; during the 
20-month episode there were about ten excursions of $\pm 25$\%
on time scales of about one month. At 8~GHz there are 
10\% variations on times of 6 months and 
during the 2~GHz episode some signs of similar
variations at a lower amplitude.  The higher
amplitudes at 2~GHz strongly suggest ISS as the cause of the 
2~GHz episode.  Is this due to a change in the source 
or the medium?  While we cannot give a definitive answer,
the stochastic appearance and slower than normal 
timescale (at a latitude of 49\de) suggests that an
unusual region of enhanced turbulence intervened 
during 1989.9--1991.6, causing strong refractive variations.
The 20-month duration at 30 km~s$^{-1}$ implies a size of 
the region $\sim 10$ AU.  

It is possible that this episode is similar to an ESE, 
but that the ISM structure
was larger including multiple regions with sufficient
refracting power to focus or defocus the 2~GHz radiation. 
Hence we looked at all sources for non-stationary
episodes and found unusual 2~GHz variations in 
0355$+$508, 0537$-$158, 0836$+$710,
1100$+$772, 1502$+$106, 1741$-$038.

\subsubsection*{1741$-$038} 

Hjellming and Narayan (1986) analyzed one year of observations 
at 1.5, 5, 15 and 22 GHz of this quasar, which
lies at a low latitude (13\de) toward the inner Galaxy. 
They identified 20-d variations at 1.5~GHz as refractive ISS 
(possibly caused at a few hundred parsecs in the North Polar Spur), 
while the higher frequencies were dominated by intrinsic variation.
Our 2~GHz results also give an 18-d timescale at an rms amplitude
of 10\%.  However, the 2~GHz auto-correlations are strongly influenced
by the inclusion of data from a well studied ESE 
(1992.2-1992.6, see Lazio et al., 2000).  
During the event the 2~GHz flux density
decreased to a minimum 50\% of normal, remaining lower 
than 75\%  for about 2 months.  

Examination of the entire 2~GHz time series reveals 
$\pm 10$\%  variations on similar timescales
from 1983.5--1994.5. There is also a period 1988.0--1989.6
during which the timescale appeared to be much shorter
($\sim 12$ d).  The VLBI observations following the minimum
in the light curve at 1992.4 showed that the source
had a diameter of 0.9 mas during the minimum flux density and 
0.6 mas afterwards.  Lazio et al.\ (2000) concluded
that the increase in diameter during the event was due to
an extra stochastic broadening.  

For refractive ISS we expect
a timescale $\sim L \theta / 1.7 V$ for a scattering 
at distance $L$ of a source observed to have a diameter $\theta$
with transverse velocity $V$.
For 0.9 (0.6) mas at 1 kpc and 30 km~s$^{-1}$ we 
obtain 32 (21) d.  These are within 50\% of our
estimated ISS timsecale and suggests that the 
conditions for the normal
ISS and the ESE are fairly similar.  
However, as noted for several of the sources discussed
in this section, there is good evidence for non-stationary
behavior in the ISS phenomenon, which is presumably based
in distinct physical structures crossing the line of sight
on occasion.
 
\subsubsection*{2005$-$403}

This quasar lies at 4 degrees latitude in the Cygnus region
and is substantially broadened by interstellar scattering. 
As shown by Desai \& Fey (2001) its scattering disc
has a diameter of 18 mas at 2.3 GHz with an axial ratio 
of (0.76:1), and its diameter scales as (wavlength)$^{2}$.
The associated refractive ISS should have a time scale
of about 1.1 years and cannot be distinguished from the slow
intrinsic variation and
(spurious) annual variation visible for this source.
However, at 8 GHz there is a 4\% variation on a timescale
of $\sim 12$ d, which we suggest is ISS. Extrapolating their
diameter measurements to 8.3 GHz gives a scattered diameter 
of $\theta_{mas}~1.3$ mas.  If the intrinsic source size was no larger 
than this we would expect the ISS timescale to 
be about 18~days~$\times L_{\rm kpc} \theta_{\rm mas}/(V_{\rm kms}/50)$,
where the scattering path is $L_{\rm kpc}
$ and $V_{\rm kms}
$
is the transverse velocity of the medium relative
to the Earth. The nominal scattering distance is 5.5 kpc, 
which would only agree with the 12 d timescale
if $V_{\rm kms} \sim 500$ km~s$^{-1}$.  This is unreasonably high
given a 220 km~s$^{-1}$ Galactic rotation velocity,
and suggests that the effective distance to the scattering
plasma is much less than 5.5 kpc. At 1 kpc the timescale
implies a transverse velocity of 90 km~s$^{-1}$ ---
still higher than the 30-50 km~s$^{-1}$ commonly assumed.
However, these figures are necessarily crude, since they do not 
include effects of anisotropic scattering or the distribution
of scattering along the path. We also note that the time-series of 8~GHz
flux density has several intervals with quasi-oscillatory
variations which have timescales noticeably longer than the 12~d
derived from the correlation analysis.

\section{Discussion and Conclusion}
\label{sec:conclusion}

We have used a simple filter on the time series of
flux density to separate slow intrinsic variations from fast
ISS on all of the light curves from the GBI monitoring
program.  The results confirm that both processes
are widespread and must be considered together in
a full description of compact radio sources.  Both processes
imply small diameters for the sources.

Our model for the interstellar scattering is that it is caused by
``turbulence'' in the ionized ISM as described by the CL05 model.
We also computed the ISS based on the TC93 model which predicted
generally slightly higher scinillation indices and about 15\% shorter
time scales. This was particularly evident in the form of the predicted
sptructure function at small time lags.

\begin{itemize}

\item We find that a simple filtering can separate the fast 
ISS from the slow intrinsic variations in flux density
at 2~GHz. A demarcation timescale of about 365~d is satisfactory
for most of the sources. 

\item We identify the 5--50~d variation in the 2~GHz flux
density of about 121 sources as ISS.  These variations are the expected
ISS phenomenon intermediate between LFV and IDV.

\item The modulation indices and timescales show clear dependence
on Galactic latitude.
This confirms that ISS is the origin of such fluctuations.
In addition there is a strong correlation of the 2~GHz scintillation index
with the emission measure estimated from WHAM H$_{\alpha}$ emission.

\item Over the 121 sources the 2~GHz ISS modulation index is
correlated with the mean spectral index and also with the 
detection of compact structure on VLBI scales.

\item A model for ISS in the CL05 (or TC93) medium is reasonably successful
if the peak brightness of the compact core of each source
is in the range $10^{11}$K to $10^{12}$K and 10-50\% of the 
source's flux density is in such a compact component.
This is in general agreement with the distribution
of brightness temperatures proposed by Readhead (1994).

\item There is a weak inverse correlation of scintillation index with
mean flux density, as expected if the effective diameter
of a source is limited by a maximum in brightness temparature.

\item  There is a significant population of sources that show
substantial 2~GHz ISS, but do not vary intrinsically 
(above about 4\% on times of a year or more).
This implies that some very compact sources do not vary
intrinsically. Conversely, there are a few 8~GHz sources with 
strong intrinsic variations with surprisingly low levels of ISS
at 2~GHz. 

\item An annual modulation in the ISS time-scale is found for
a few sources, confirming both the ISS model
and that in these directions the scattering 
plasma moves approximately with the local standard
of rest.  However, most sources do not show such a simple annual
cycle, implying that we do not have an accurate understanding
of the velocity of the scattering plasma.

\item An intrinsic outburst from the BL Lac object 0235$+$164 was detected
in 1992.7 by the sudden onset of ISS at 2~GHz while its total flux density
started a steep rise.  We estimated the brief emergence
of a component as bright as $1.7 \times 10^{14}$K which
decreased over a few days as it expanded, eventually doubling the total
flux density of the source.

\item At Galactic latitudes less than about 15 degrees
from the plane the ISS model needs to include
differential Galactic rotation, which will increase the  
effective velocity of the ISM relative to the model and
so decrease the predicted time-scale,  This may include
a wide enough velocity range along the line of sight
to invalidate the ``frozen pattern'' assumption.

\end{itemize}

\acknowledgements
We thank A.~Fey for helpful discussions and for providing
the VLBI data.  We thank E.~Waltman for her many years of work with the
\hbox{GBI}, without which both the data quality and quantity
would have been much reduced.
The Green Bank Interferometer is a facility of the National Science
Foundation operated by the National Radio Astronomy Observatory, 
under contract to the US Naval
Observatory and Naval Research Laboratory (NRL) in support of their
geodetic and astronomy programs. 
Basic research in radio astronomy at the NRL is supported by the
Office of Naval Research. The National Radio 
Astronomy Observatory is a facility of
the National Science Foundation operated under cooperative
agreement by Associated Universities, Inc.
The interstellar research at UCSD was partially funded by
the National Science Foundation under grants AST 9988398 and 0507713. 
We thank Bill Coles for valuable discussions and we
thank Ron Reynolds for access to the data from the Wisconsin H-Alpha Mapper,
which is funded by the National Science Foundation.

\appendix
\section{Intensity correlation for WISS and RISS}\label{app:iss}

In the ISS model of Rickett et al. (1995) a Gaussian 
function was assumed to describe the dependence of $C_N^2$ on distance from the
Earth; here we use the same basic method, but instead integrate the CL05 or TC93 model in the direction toward each source. 

For a source of unit flux density the intensity covariance function
(at offset ${\bm s}$) is given by CFRC equation (11), 
which we rearrange in terms of distance $z$ measured
from the observer:
\begin{eqnarray}
C_{\delta I}({\bm s}) = 8 \pi r_e^2  \lambda^2  \int _{0}^{\infty} dz
\int _{0}^{\infty} d^2{\bm q} C_N^2(z) q^{-\alpha -2} \times \nonumber\\
|V(qz/k)|^2 \; \sin^2 [\frac{q^2 \lambda z}{4\pi}] \;\exp[ -D_r(q,z) + i{\bm q \cdot s} ] 
\label{eq:cdeltaI} 
\end{eqnarray}
Here $k=2\pi/\lambda$ for wavelength $\lambda$, $r_e$ is the classical electron radius,
$\alpha = 5/3$ for the assumed Kolmogorov spectrum of density as
$ C_N^2(z) q^{-\alpha -2}$ at transverse wavenumber $q$.  The refractive
cut-off is given by:
\be
D_r(q,z)
 = \int _{0}^{\infty} dt \; D'(q \; {\rm min}(t,z)/k,t)  = (qL_{\rm eff}(z) \theta_s)^{\alpha} 
\label{eq:dr} 
\ee
and $D'(s,z)$ is given by equation (4) of CFRC.  The second part
of equation \ref{eq:dr} involves an effective path length through the scattering
medium:
\begin{eqnarray}
L_{\rm eff}(z)^{\alpha} = \int _{0}^{\infty} dt \; [{\rm min}(t,z)]^{\alpha} \; C_N^2(t)\; /SM \nonumber \\
SM = \int _{0}^{\infty} dt \; C_N^2(t)
\label{eq:leff}
\end{eqnarray}
It follows that near the observer the scattering distance is effectively 
$L_{\rm eff}(z) \sim z$; but at large distances from the observer
$L_{\rm eff}$ saturates at about the 1/e scale of the 
$C_N^2$ distribution. The scattering angle~$\theta_s$ 
corresponds to the effective angular broadening at the observer:
\be
\theta_s = k^{-1}\left [ 8 \pi^2 r_e^2  \lambda^2 SM 
\frac{\Gamma (1-\alpha/2)}{\Gamma (1+\alpha/2) \alpha 2^{\alpha}} \right ]^{-1/\alpha}
\ee

We proceed as in CFRC to approximate the exponent $\alpha$ in
\ref{eq:dr} by a value two in place of 5/3.  As this only governs the precise
form of the refractive cut-off it changes the result very little.
We further assume that the source visibility is a circular
Gaussian function:
\be
V(u) = \exp [-0.5(u k \theta_{so})^2]
\ee
in which the source diameter is 2.35 $\theta_{so}$.
These two assumptions make a Gaussian cut-off in the wavenumber integral,
which allows the integral to be done as in the Appendix of CFRC
in terms of a confluent hypergeometric function.
We then numerically integrate equation \ref{eq:cdeltaI} with the TC93 weighting for
$C_N^2$ over z to obtain $C_{\delta I}({\bm s})$.

The modulation index is then given by $m = \sqrt{C_{\delta I}(0)}$
and the spatial scale $s_{iss}$ is found by solving for
where the correlation falls to 0.5 of its maximum: 
$C_{\delta I}({\bm s_{iss}}) = 0.5 C_{\delta I}(0)$.

\clearpage

\begin{deluxetable}{rrrrrrrrrrrrrrrcc} 
\tabletypesize{\scriptsize}
\rotate
\tablewidth{0pc} 
\tablecaption{Short-term Variations in GBI Flux Density Monitoring}
\tablehead{\colhead{Name} &  \colhead{$l$}  &  \colhead{$b$}  & 
\colhead{$L_{\rm eff}$}  &  \colhead{$S_2$} &   \colhead{$S_8$}  &
\colhead{$m_2$(rel)}  &  \colhead{$\tau_2$} & $T_{\rm sm}$ &
\colhead{$m_8$(rel)}    &  \colhead{$\tau_8$} &
\colhead{$M_2$} & \colhead{$M_8$} & 
\colhead{$\rho$} &  \colhead{spectral index}  & \colhead{$S_{2,vlbi}$}  & 
\colhead{Diameter} \\ 
\colhead{1950} & \colhead{deg} & \colhead{deg} &  \colhead{kpc} &
\colhead{Jy}  &  \colhead{Jy} &   &  \colhead{days} &  \colhead{days} &   &
\colhead{days} &  &  &  & \colhead{$\alpha$} & \colhead{Jy} & \colhead{mas}}    
\startdata
0003$+$380 & 113 & -24 & 1.61  &  0.59 &  0.89  & 0.048(2)  &   8.0 &  365  & 0.044(1)  &  28.2  & 0.035 & 0.126 &  0.23 &  0.32 &  0.61 &  1.25\\ 
0003$-$066 &  93 & -67 & 0.68  &  1.98 &  2.22  & 0.015(2)  &   6.3 &  365  & 0.023(0)  &  19.7  & 0.079 & 0.091 &   -   &  0.09 &  2.27 &  1.21\\ 
0016$+$731 & 121 &  11 & 2.55  &  1.39 &  1.64  & 0.044(2)  &  21.9 &  365  & 0.025(1)  &   8.6  & 0.053 & 0.042 &  0.14 &  0.13 &  1.53 &  1.23\\ 
0019$+$058 & 110 & -56 & 0.77  &  0.42 &  0.44  & 0.031(1)  &   4.6 &  200  & 0.034(1)  &  10.0  & 0.269 & 0.399 &   -   &  0.03 &  0.25 &  0.61\\ 
0035$+$121 & 118 & -50 & 0.83  &  0.77 &  0.38  & 0.007(0)  &   6.2 &  365  & 0.014(0)  &   8.2  & 0.008 & 0.039 &   -   & -0.53 &  0.00 &  0.00\\ 
0035$+$413 & 120 & -21 & 1.67  &  0.97 &  1.01  & 0.030(2)  &  10.0 &  365  & 0.014(1)  &   8.6  & 0.025 & 0.036 &  0.25 &  0.03 &  0.00 &  0.00\\ 
0055$+$300 & 125 & -33 & 1.17  &  0.42 &  0.68  & 0.012(2)  &  12.1 &  365  & 0.016(1)  &  12.7  & 0.017 & 0.044 &  0.79 &  0.37 &  0.00 &  0.00\\ 
0056$-$001 & 127 & -63 & 0.68  &  2.00 &  0.87  & 0.007(1)  &   4.8 &  365  & 0.012(0)  &  17.5  & 0.008 & 0.032 &   -   & -0.63 &  0.32 &  3.75\\ 
0113$-$118 & 144 & -73 & 0.62  &  1.62 &  1.48  & 0.009(1)  &   5.4 &  365  & 0.013(0)  &  20.2  & 0.053 & 0.107 &   -   & -0.06 &  0.86 &  2.69\\ 
0123$+$257 & 133 & -36 & 1.03  &  1.14 &  0.95  & 0.012(1)  &  10.5 &  365  & 0.013(0)  &  10.4  & 0.036 & 0.043 &   -   & -0.14 &  0.87 &  0.75\\ 
0130$-$171 & 168 & -76 & 0.62  &  0.77 &  0.59  & 0.013(1)  &   5.0 &  365  & 0.019(0)  &   7.6  & 0.030 & 0.072 &   -   & -0.20 &  0.53 &  1.51\\ 
0133$+$476 & 131 & -14 & 2.04  &  1.43 &  1.60  & 0.049(2)  &   8.6 &  365  & 0.031(1)  &   5.8  & 0.068 & 0.102 &  0.18 &  0.20 &  1.43 &  2.25\\ 
0134$+$329 & 134 & -29 & 1.23  & 11.35 &  3.33  & 0.006(1)  &   9.6 &  365  & 0.012(0)  &   3.8  & 0.005 & 0.009 &   -   & -0.94 &  0.23 &  1.17\\ 
0147$+$187 & 142 & -42 & 0.88  &  0.44 &  0.45  & 0.022(1)  &   4.8 &  365  & 0.017(0)  &  18.3  & 0.053 & 0.102 &   -   &  0.02 &  0.00 &  0.00\\ 
0201$+$113 & 150 & -48 & 0.78  &  0.91 &  0.73  & 0.014(2)  &   5.0 &  365  & 0.018(1)  &  18.1  & 0.010 & 0.099 &  0.14 & -0.17 &  0.92 &  1.00\\ 
0202$+$319 & 141 & -28 & 1.20  &  0.75 &  0.71  & 0.020(2)  &   6.4 &  365  & 0.021(0)  &   9.6  & 0.034 & 0.050 &   -   & -0.02 &  0.35 &  0.01\\ 
0212$+$735 & 129 &  12 & 2.27  &  2.38 &  2.91  & 0.015(2)  &  11.2 &  365  & 0.019(1)  &   8.6  & 0.015 & 0.071 &  0.25 &  0.16 &  1.57 &  1.46\\ 
0224$+$671 & 132 &   6 & 2.52  &  1.59 &  2.11  & 0.032(2)  &  11.2 &  200  & 0.036(1)  &   9.9  & 0.124 & 0.186 &  0.34 &  0.16 &  1.21 &  0.46\\ 
0235$+$164 & 157 & -39 & 0.87  &  1.75 &  2.11  & 0.017(1)  &   2.9 &  100  & 0.021(0)  &   5.6  & 0.331 & 0.419 &   -   &  0.09 &  0.72 &  1.06\\ 
0237$-$233 &  30 & -65 & 0.66  &  4.82 &  1.72  & 0.005(1)  &   3.4 &  365  & 0.018(1)  &   5.0  & 0.004 & 0.027 &   -   & -0.67 &  3.91 &  1.55\\ 
0256$+$075 & 169 & -44 & 0.78  &  0.75 &  0.61  & 0.029(2)  &   7.0 &  365  & 0.037(0)  &  22.0  & 0.081 & 0.161 &   -   & -0.17 &  0.72 &  0.65\\ 
0300$+$470 & 145 & -10 & 2.13  &  1.43 &  1.51  & 0.066(2)  &  19.4 &  365  & 0.042(1)  &  10.5  & 0.085 & 0.088 &  0.25 &  0.12 &  0.95 &  0.47\\ 
0316$+$413 & 151 & -13 & 1.86  & 33.49 & 30.33  & 0.007(1)  &  15.6 &  365  & 0.017(1)  &   5.8  & 0.017 & 0.028 &   -   & -0.08 &  0.00 &  0.00\\ 
0319$+$121 & 171 & -36 & 0.87  &  1.65 &  1.12  & 0.014(2)  &   7.2 &  365  & 0.013(1)  &   4.3  & 0.024 & 0.025 &  0.67 & -0.30 &  0.94 &  3.48\\ 
0333$+$321 & 159 & -19 & 1.45  &  2.51 &  1.64  & 0.031(2)  &  14.5 &  365  & 0.027(1)  &   6.0  & 0.019 & 0.056 &  0.28 & -0.39 &  0.55 &  1.75\\ 
0336$-$019 &   8 & -43 & 0.89  &  2.43 &  2.23  & 0.043(2)  &   5.3 &  365  & 0.024(0)  &  22.6  & 0.056 & 0.087 &   -   & -0.02 &  2.27 &  1.58\\ 
0337$+$319 & 160 & -18 & 1.46  &  0.22 &  0.03  & 0.017(1)  &  15.6 &  365  & 0.040(0)  &   6.7  & 0.031 & 0.060 &   -   & -1.43 &  0.00 &  0.00\\ 
0355$+$508 & 150 &  -2 & 2.30  &  4.07 &  2.57  & 0.011(1)  &  24.6 &  365  & 0.034(1)  &  13.6  & 0.043 & 0.068 &   -   & -0.07 &  1.54 &  4.97\\ 
0400$+$258 & 168 & -20 & 1.34  &  0.99 &  0.77  & 0.016(2)  &   5.6 &  365  & 0.012(1)  &  13.9  & 0.022 & 0.032 &  0.36 & -0.20 &  0.60 &  0.52\\ 
0403$-$132 & 206 & -43 & 0.92  &  3.24 &  1.65  & 0.012(1)  &   5.3 &  365  & 0.021(0)  &  20.0  & 0.016 & 0.050 &   -   & -0.52 &  0.44 &  0.78\\ 
0420$-$014 &  15 & -33 & 1.10  &  3.55 &  3.50  & 0.030(2)  &   7.3 &  365  & 0.024(1)  &   9.6  & 0.069 & 0.099 &  0.11 &  0.13 &  3.10 &  0.74\\ 
0440$-$003 & 197 & -28 & 1.18  &  1.54 &  0.93  & 0.039(2)  &   6.4 &  365  & 0.042(1)  &  28.5  & 0.034 & 0.151 &  0.14 & -0.38 &  2.11 &  1.16\\ 
0444$+$634 & 146 &  12 & 2.00  &  0.43 &  0.59  & 0.065(2)  &  13.8 &  365  & 0.029(0)  &   5.9  & 0.116 & 0.196 &   -   &  0.24 &  0.00 &  0.00\\ 
0454$+$844 & 128 &  25 & 1.43  &  0.42 &  0.43  & 0.038(1)  &   5.1 &  365  & 0.026(1)  &  11.2  & 0.050 & 0.087 &   -   &  0.02 &  0.39 &  0.70\\ 
0500$+$019 & 198 & -23 & 1.37  &  2.42 &  1.46  & 0.012(2)  &   7.2 &  365  & 0.010(1)  &   8.9  & 0.011 & 0.030 &  0.49 & -0.38 &  0.00 &  0.00\\ 
0528$+$134 & 191 & -11 & 1.98  &  2.27 &  3.46  & 0.055(2)  &  12.4 &  365  & 0.033(1)  &   5.4  & 0.054 & 0.240 &  0.26 &  0.32 &  2.38 &  1.82\\ 
0532$+$826 & 131 &  25 & 1.40  &  0.26 &  0.22  & 0.029(2)  &   5.4 &  365  & 0.034(1)  &  24.0  & 0.044 & 0.104 &  0.26 & -0.12 &  0.00 &  0.00\\ 
0537$-$158 & 220 & -23 & 1.68  &  0.47 &  0.30  & 0.033(2)  &  12.9 &  365  & 0.012(0)  &   9.4  & 0.019 & 0.057 &   -   & -0.34 &  0.37 &  1.62\\ 
0538$+$498 & 162 &  10 & 1.96  & 15.58 &  4.89  & 0.008(1)  &  18.0 &  365  & 0.013(0)  &  11.2  & 0.009 & 0.013 &   -   & -0.89 &  1.64 &  2.75\\ 
0552$+$398 & 172 &   7 & 2.04  &  3.29 &  7.03  & 0.013(2)  &  23.2 &  365  & 0.015(0)  &   9.9  & 0.013 & 0.048 &   -   &  0.40 &  4.00 &  1.38\\ 
0555$-$132 & 219 & -18 & 2.02  &  0.76 &  0.64  & 0.029(2)  &  10.1 &  365  & 0.014(0)  &   4.0  & 0.018 & 0.097 &   -   & -0.13 &  0.00 &  0.00\\ 
0615$+$820 & 132 &  26 & 1.34  &  0.83 &  0.67  & 0.021(2)  &   8.4 &  365  & 0.018(1)  &  10.5  & 0.040 & 0.062 &  0.36 & -0.17 &  0.91 &  0.90\\ 
0624$-$058 & 215 &  -8 & 2.79  &  9.58 &  2.67  & 0.007(1)  &  24.5 &  365  & 0.010(0)  &   7.8  & 0.005 & 0.029 &   -   & -0.98 &  0.00 &  0.00\\ 
0633$+$734 & 141 &  25 & 1.30  &  0.92 &  0.83  & 0.021(2)  &   7.1 &  365  & 0.018(1)  &  17.6  & 0.011 & 0.062 &  0.57 & -0.08 &  0.00 &  0.00\\ 
0650$+$371 & 179 &  16 & 1.48  &  1.03 &  0.83  & 0.020(2)  &   5.9 &  365  & 0.012(0)  &   3.3  & 0.045 & 0.053 &   -   & -0.16 &  0.65 &  0.82\\ 
0653$-$033 & 216 &  -1 & 2.94  &  0.49 &  0.52  & 0.065(2)  &  16.0 &  365  & 0.039(1)  &   8.8  & 0.092 & 0.107 &  0.05 &  0.05 &  0.00 &  0.00\\ 
0716$+$714 & 144 &  28 & 1.18  &  0.44 &  0.60  & 0.046(1)  &   5.1 &  365  & 0.033(1)  &   6.7  & 0.102 & 0.248 &   -   &  0.27 &  0.27 &  0.51\\ 
0723$+$679 & 148 &  28 & 1.14  &  0.71 &  0.36  & 0.012(1)  &  10.5 &  365  & 0.022(1)  &   9.0  & 0.017 & 0.055 &   -   & -0.53 &  0.00 &  0.00\\ 
0723$-$008 & 218 &   7 & 2.87  &  1.54 &  1.20  & 0.036(2)  &   4.2 &  365  & 0.018(1)  &  28.9  & 0.030 & 0.066 &  0.44 & -0.19 &  0.99 &  1.26\\ 
0742$+$103 &  30 &  16 & 2.23  &  4.21 &  2.58  & 0.011(1)  &   5.3 &  365  & 0.013(0)  &  10.0  & 0.007 & 0.032 &   -   & -0.32 &  3.85 &  1.39\\ 
0743$+$259 & 195 &  23 & 1.22  &  0.49 &  0.39  & 0.026(1)  &   5.6 &  365  & 0.029(0)  &  17.2  & 0.044 & 0.220 &   -   & -0.18 &  0.46 &  0.52\\ 
0759$+$183 & 204 &  23 & 1.27  &  0.51 &  0.57  & 0.042(2)  &   6.5 &  365  & 0.005(0)  &  15.9  & 0.019 & 0.053 &   -   &  0.09 &  0.00 &  0.00\\ 
0804$+$499 & 169 &  32 & 0.94  &  1.10 &  1.25  & 0.022(1)  &   4.7 &  365  & 0.000(0)  & -46.0  & 0.102 & 0.276 &   -   &  0.10 &  1.15 &  0.99\\ 
0818$-$128 & 235 &  13 & 2.57  &  0.85 &  0.60  & 0.045(2)  &   6.3 &  365  & 0.024(1)  &  18.2  & 0.049 & 0.084 &  0.58 & -0.26 &  0.41 &  0.95\\ 
0827$+$243 & 200 &  32 & 0.94  &  0.65 &  0.79  & 0.025(2)  &  10.4 &  365  & 0.033(0)  &  29.5  & 0.068 & 0.243 &   -   &  0.14 &  0.46 &  0.75\\ 
0836$+$710 & 144 &  34 & 1.01  &  3.14 &  1.83  & 0.030(2)  &  14.0 &  365  & 0.038(1)  &  20.1  & 0.049 & 0.107 &  0.35 & -0.41 &  1.33 &  0.93\\ 
0837$+$035 & 223 &  25 & 1.36  &  0.73 &  0.56  & 0.017(2)  &   5.5 &  365  & 0.004(0)  &   3.4  & 0.009 & 0.031 &   -   & -0.21 &  0.00 &  0.00\\ 
0851$+$202 &  27 &  36 & 1.24  &  2.24 &  3.12  & 0.023(1)  &   6.0 &  150  & 0.029(1)  &   6.5  & 0.144 & 0.229 &   -   &  0.30 &  1.40 &  1.42\\ 
0859$-$140 & 242 &  21 & 1.77  &  2.48 &  1.51  & 0.011(1)  &  23.1 &  365  & 0.011(0)  &   6.7  & 0.007 & 0.030 &   -   & -0.38 &  1.77 &  1.46\\ 
0922$+$005 & 232 &  34 & 1.08  &  0.82 &  0.53  & 0.031(1)  &   9.4 &  365  & 0.009(0)  &   9.0  & 0.036 & 0.057 &   -   & -0.33 &  0.00 &  0.00\\ 
0923$+$392 &   4 &  46 & 1.07  &  4.66 & 10.37  & 0.010(1)  &  13.7 &  365  & 0.016(0)  &  25.1  & 0.009 & 0.058 &   -   &  0.39 &  4.97 &  1.22\\ 
0938$+$119 & 222 &  43 & 0.84  &  0.19 &  0.08  & 0.010(1)  &  14.4 &  365  & 0.028(0)  &  45.3  & 0.010 & 0.099 &   -   & -0.65 &  0.00 &  0.00\\ 
0945$+$408 & 181 &  50 & 0.68  &  1.45 &  1.62  & 0.019(2)  &   7.2 &  365  & 0.016(1)  &   6.4  & 0.040 & 0.092 &  0.64 &  0.08 &  1.25 &  0.35\\ 
0952$+$179 & 217 &  48 & 0.74  &  1.01 &  0.54  & 0.009(1)  &   6.5 &  365  & 0.011(0)  &  23.0  & 0.006 & 0.035 &   -   & -0.47 &  0.45 &  1.37\\ 
0954$+$658 & 146 &  43 & 0.85  &  0.92 &  0.99  & 0.034(2)  &   3.9 &  150  & 0.040(1)  &   4.4  & 0.108 & 0.133 &  0.49 &  0.10 &  0.67 &  1.13\\ 
1020$+$400 & 181 &  57 & 0.65  &  0.64 &  1.05  & 0.014(1)  &   7.2 &  365  & 0.011(0)  &  24.7  & 0.063 & 0.107 &   -   &  0.38 &  0.61 &  0.45\\ 
1022$+$194 & 218 &  55 & 0.67  &  0.44 &  0.64  & 0.016(1)  &   6.6 &  365  & 0.011(0)  &   8.0  & 0.016 & 0.031 &   -   &  0.30 &  0.00 &  0.00\\ 
1036$-$154 & 262 &  36 & 1.03  &  0.43 &  0.45  & 0.023(1)  &   5.0 &  365  & 0.023(1)  &  21.4  & 0.057 & 0.122 &   -   &  0.04 &  0.00 &  0.00\\ 
1038$+$528 & 158 &  55 & 0.69  &  0.29 &  0.71  & 0.032(1)  &   5.5 &  365  & 0.017(0)  &  12.6  & 0.072 & 0.127 &   -   &  0.68 &  0.00 &  0.00\\ 
1055$+$018 & 251 &  53 & 0.73  &  2.75 &  3.30  & 0.010(1)  &   6.1 &  365  & 0.015(0)  &  23.6  & 0.058 & 0.067 &   -   &  0.14 &  1.31 &  2.80\\ 
1100$+$772 & 130 &  39 & 0.99  &  0.38 &  0.19  & 0.014(2)  &  13.4 &  365  & 0.030(0)  &   5.6  & 0.009 & 0.063 &   -   & -0.53 &  0.00 &  0.00\\ 
1116$+$128 & 242 &  64 & 0.67  &  1.98 &  1.46  & 0.009(1)  &  13.2 &  365  & 0.013(0)  &   7.7  & 0.018 & 0.070 &   -   & -0.23 &  1.05 &  0.85\\ 
1123$+$264 & 211 &  71 & 0.74  &  1.10 &  1.05  & 0.011(1)  &   8.2 &  365  & 0.010(0)  &   5.8  & 0.021 & 0.056 &   -   & -0.03 &  1.14 &  1.17\\ 
1127$-$145 & 275 &  44 & 0.86  &  4.93 &  2.55  & 0.007(1)  &  12.0 &  365  & 0.015(0)  &   7.9  & 0.018 & 0.062 &   -   & -0.50 &  3.27 &  0.95\\ 
1128$+$385 & 175 &  70 & 0.89  &  0.78 &  0.93  & 0.019(1)  &   7.4 &  365  & 0.017(0)  &  25.9  & 0.012 & 0.088 &   -   &  0.13 &  0.64 &  0.80\\ 
1145$-$071 & 277 &  52 & 0.77  &  0.89 &  0.76  & 0.009(1)  &   5.9 &  365  & 0.014(0)  &  15.5  & 0.012 & 0.093 &   -   & -0.12 &  0.78 &  1.57\\ 
1150$+$812 & 126 &  36 & 1.08  &  1.30 &  1.42  & 0.016(1)  &   7.5 &  365  & 0.021(1)  &  16.0  & 0.024 & 0.062 &   -   &  0.07 &  1.11 &  0.53\\ 
1155$+$251 &  41 &  78 & 0.80  &  1.22 &  0.70  & 0.005(1)  &  12.9 &  365  & 0.014(0)  &   7.2  & 0.016 & 0.027 &   -   & -0.42 &  0.62 &  3.00\\ 
1200$-$051 & 281 &  55 & 0.76  &  0.41 &  0.50  & 0.015(1)  &   7.0 &  365  & 0.023(0)  &  33.8  & 0.058 & 0.188 &   -   &  0.15 &  0.00 &  0.00\\ 
1225$+$368 & 150 &  79 & 0.86  &  1.77 &  0.39  & 0.006(1)  &  18.6 &  365  & 0.011(0)  &   6.5  & 0.005 & 0.013 &   -   & -1.16 &  0.00 &  0.00\\ 
1243$-$072 & 300 &  55 & 0.77  &  0.61 &  0.72  & 0.035(2)  &  12.8 &  365  & 0.014(1)  &  13.9  & 0.024 & 0.084 &  0.29 &  0.12 &  0.52 &  1.26\\ 
1245$-$197 & 122 &  43 & 0.94  &  3.74 &  1.27  & 0.003(2)  &  12.7 &  365  & 0.013(0)  &   3.3  & 0.001 & 0.019 &   -   & -0.74 &  0.00 &  0.00\\ 
1250$+$568 & 123 &  61 & 0.71  &  1.62 &  0.45  & 0.007(1)  &  10.7 &  365  & 0.018(0)  &   4.9  & 0.009 & 0.033 &   -   & -0.98 &  0.00 &  0.00\\ 
1253$-$055 & 305 &  57 & 0.77  &  9.32 & 10.89  & 0.008(2)  &   8.9 &  365  & 0.015(0)  &  32.1  & 0.020 & 0.070 &   -   &  0.12 &  7.76 &  1.75\\ 
1302$-$102 & 308 &  52 & 0.80  &  0.80 &  0.72  & 0.020(2)  &   7.2 &  365  & 0.023(0)  &  36.1  & 0.080 & 0.109 &   -   & -0.08 &  1.04 &  0.52\\ 
1308$+$326 &  86 &  83 & 0.84  &  1.35 &  2.89  & 0.018(1)  &   4.0 &  365  & 0.016(0)  &  11.9  & 0.128 & 0.136 &   -   &  0.58 &  2.32 &  0.25\\ 
1328$+$254 &  22 &  81 & 0.75  &  5.48 &  2.20  & 0.003(1)  &   9.8 &  365  & 0.009(0)  &   2.4  & 0.003 & 0.008 &   -   & -0.64 &  0.00 &  0.00\\ 
1328$+$307 &  56 &  81 & 0.77  & 11.88 &  5.36  & 0.005(1)  &   2.7 &  365  & 0.015(0)  &   3.1  & 0.005 & 0.010 &   -   & -0.57 &  0.00 &  0.00\\ 
1354$+$195 &   9 &  73 & 0.85  &  1.80 &  1.24  & 0.011(1)  &   8.6 &  365  & 0.006(0)  &  10.0  & 0.021 & 0.057 &   -   & -0.29 &  0.00 &  0.00\\ 
1404$+$286 &  42 &  73 & 0.85  &  1.69 &  1.93  & 0.005(1)  &  13.0 &  365  & 0.012(0)  &  12.8  & 0.006 & 0.020 &   -   &  0.12 &  1.96 &  1.62\\ 
1409$+$524 &  98 &  61 & 0.73  & 14.06 &  2.07  & 0.005(1)  &  10.3 &  365  & 0.018(1)  &  15.9  & 0.006 & 0.035 &   -   & -1.46 &  0.00 &  0.00\\ 
1413$+$135 &   2 &  66 & 0.90  &  0.85 &  1.31  & 0.005(1)  &   7.8 &  200  & 0.026(0)  &   9.0  & 0.016 & 0.230 &   -   &  0.33 &  0.26 &  1.90\\ 
1430$-$155 & 336 &  41 & 1.06  &  0.64 &  0.72  & 0.034(2)  &   6.9 &  365  & 0.014(0)  &   6.4  & 0.009 & 0.037 &   -   &  0.08 &  0.00 &  0.00\\ 
1438$+$385 &  66 &  65 & 0.90  &  0.90 &  0.41  & 0.008(1)  &  10.6 &  365  & 0.015(0)  &  17.1  & 0.010 & 0.030 &   -   & -0.59 &  0.00 &  0.00\\ 
1449$-$012 & 353 &  49 & 1.01  &  0.49 &  0.34  & 0.015(1)  &   6.8 &  365  & 0.016(0)  &   4.9  & 0.011 & 0.079 &   -   & -0.28 &  0.00 &  0.00\\ 
1455$+$247 &  35 &  62 & 0.99  &  0.62 &  0.26  & 0.004(0)  &  11.6 &  365  & 0.013(0)  &   3.0  & 0.004 & 0.014 &   -   & -0.66 &  0.00 &  0.00\\ 
1502$+$106 &  11 &  55 & 1.02  &  1.89 &  1.55  & 0.028(2)  &  10.0 &  200  & 0.016(0)  &   2.8  & 0.070 & 0.114 &   -   & -0.11 &  1.64 &  0.34\\ 
1511$+$238 &  35 &  58 & 1.02  &  1.37 &  0.54  & 0.004(0)  &  14.3 &  365  & 0.011(0)  &  23.6  & 0.005 & 0.018 &   -   & -0.65 &  0.00 &  0.00\\ 
1514$+$197 &  28 &  56 & 1.03  &  0.54 &  0.68  & 0.075(2)  &   6.7 &  365  & 0.021(1)  &  11.5  & 0.089 & 0.141 &  0.49 &  0.18 &  0.47 &  1.83\\ 
1525$+$314 &  49 &  56 & 1.03  &  0.77 &  0.42  & 0.008(0)  &   3.2 &  365  & 0.021(0)  &  18.4  & 0.013 & 0.018 &   -   & -0.46 &  0.00 &  0.00\\ 
1538$+$149 &  24 &  49 & 1.09  &  1.26 &  0.93  & 0.043(2)  &   8.4 &  365  & 0.031(1)  &  18.6  & 0.033 & 0.052 &  0.38 & -0.24 &  0.56 &  1.48\\ 
1547$+$508 &  80 &  49 & 0.91  &  0.70 &  1.18  & 0.011(0)  &   3.6 &  365  & 0.017(0)  &   3.0  & 0.005 & 0.065 &   -   &  0.39 &  0.00 &  0.00\\ 
1555$+$001 &  10 &  38 & 1.19  &  0.97 &  1.12  & 0.036(2)  &   7.7 &  365  & 0.023(1)  &  17.1  & 0.091 & 0.094 &  0.33 &  0.04 &  1.02 &  1.05\\ 
1611$+$343 &  55 &  47 & 1.08  &  3.29 &  3.24  & 0.014(1)  &  10.1 &  365  & 0.015(1)  &  17.9  & 0.026 & 0.042 &   -   & -0.07 &  4.16 &  2.11\\ 
1614$+$051 &  18 &  36 & 1.22  &  0.63 &  0.64  & 0.033(2)  &  16.3 &  365  & 0.015(0)  &   6.0  & 0.033 & 0.053 &   -   &  0.02 &  0.68 &  1.07\\ 
1624$+$416 &  66 &  44 & 1.03  &  1.60 &  1.01  & 0.007(1)  &   8.2 &  365  & 0.012(0)  &  20.7  & 0.019 & 0.026 &   -   & -0.35 &  0.82 &  0.99\\ 
1635$-$035 &  13 &  27 & 1.42  &  0.39 &  0.33  & 0.066(2)  &  17.8 &  365  & 0.029(1)  &   7.9  & 0.047 & 0.063 &  0.56 & -0.13 &  0.00 &  0.00\\ 
1641$+$399 &  63 &  41 & 1.11  &  7.55 &  7.95  & 0.012(2)  &  14.1 &  365  & 0.014(0)  &   3.2  & 0.046 & 0.123 &   -   &  0.14 &  5.14 &  1.48\\ 
1655$+$077 &  27 &  29 & 1.40  &  1.07 &  0.97  & 0.021(2)  &  16.0 &  365  & 0.024(1)  &   6.0  & 0.056 & 0.090 &  0.54 & -0.08 &  0.68 &  2.02\\ 
1656$+$477 &  74 &  39 & 1.18  &  1.22 &  1.24  & 0.024(2)  &  13.2 &  365  & 0.011(0)  &   3.4  & 0.011 & 0.020 &   -   &  0.01 &  1.06 &  0.52\\ 
1741$-$038 &  22 &  13 & 2.45  &  2.03 &  2.59  & 0.097(2)  &  18.2 &  365  & 0.027(1)  &   7.4  & 0.070 & 0.058 &  0.19 &  0.19 &  1.77 &  0.70\\ 
1742$-$289 & 173 &   0 & 2.18  &  0.46 &  0.81  & 0.049(1)  &   4.3 &  365  & 0.056(1)  &  14.1  & 0.025 & 0.062 &   -   &  0.51 &  0.00 &  0.00\\ 
1749$+$096 &  35 &  18 & 2.19  &  1.14 &  2.40  & 0.059(2)  &   8.0 &  150  & 0.036(2)  &   5.9  & 0.155 & 0.264 &  0.21 &  0.48 &  1.48 &  0.43\\ 
1749$+$701 & 101 &  31 & 1.39  &  0.73 &  0.62  & 0.030(2)  &   9.4 &  365  & 0.033(1)  &  15.2  & 0.078 & 0.126 &  0.60 & -0.11 &  0.61 &  2.11\\ 
1756$+$237 &  49 &  22 & 2.00  &  0.81 &  0.59  & 0.032(2)  &   7.7 &  365  & 0.014(1)  &   6.2  & 0.012 & 0.029 &  0.40 & -0.24 &  0.00 &  0.00\\ 
1803$+$784 & 110 &  29 & 1.39  &  2.23 &  2.87  & 0.026(2)  &   6.5 &  365  & 0.027(1)  &  12.2  & 0.058 & 0.073 &  0.40 &  0.19 &  1.63 &  1.37\\ 
1807$+$698 & 100 &  29 & 1.46  &  1.47 &  1.75  & 0.015(2)  &  10.4 &  365  & 0.020(1)  &  12.6  & 0.020 & 0.061 &  0.65 &  0.18 &  0.79 &  1.99\\ 
1821$+$107 &  39 &  11 & 3.35  &  1.11 &  0.75  & 0.034(2)  &  15.7 &  365  & 0.021(1)  &   6.2  & 0.019 & 0.031 &  0.56 & -0.26 &  1.29 &  0.60\\ 
1823$+$568 &  86 &  26 & 1.74  &  1.18 &  1.38  & 0.029(2)  &   5.2 &  365  & 0.016(1)  &  10.0  & 0.019 & 0.091 &  0.63 &  0.12 &  0.84 &  0.61\\ 
1828$+$487 &  77 &  24 & 1.97  &  8.31 &  3.54  & 0.008(1)  &  14.1 &  365  & 0.011(0)  &  10.9  & 0.007 & 0.046 &   -   & -0.65 &  0.00 &  0.00\\ 
1830$+$285 &  57 &  17 & 2.70  &  0.47 &  0.53  & 0.046(2)  &  10.7 &  365  & 0.018(1)  &   6.5  & 0.050 & 0.068 &  0.33 &  0.08 &  0.48 &  2.43\\ 
1928$+$738 & 106 &  24 & 1.70  &  3.75 &  4.00  & 0.015(1)  &  11.4 &  365  & 0.017(0)  &   8.9  & 0.019 & 0.033 &   -   &  0.05 &  0.00 &  0.00\\ 
1943$+$228 &  59 &  -1 & 8.37  &  0.31 &  0.16  & 0.020(1)  &  21.0 &  365  & 0.031(0)  &   4.8  & 0.026 & 0.064 &   -   & -0.52 &  0.00 &  0.00\\ 
1944$+$251 &  61 &   0 & 8.51  &  0.32 &  0.10  & 0.014(1)  &  11.1 &  365  & 0.080(0)  &   4.4  & 0.015 & 0.133 &   -   & -0.89 &  0.27 &  3.05\\ 
1947$+$079 &  47 &  -9 & 4.13  &  1.30 &  0.72  & 0.021(2)  &  18.6 &  365  & 0.008(0)  &  14.0  & 0.015 & 0.028 &   -   & -0.45 &  0.00 &  0.00\\ 
2005$+$403 &  77 &   4 & 2.86  &  2.79 &  3.31  & 0.007(1)  &   3.9 &  365  & 0.037(2)  &  11.8  & 0.033 & 0.032 &   -   &  0.05 &  0.87 &  1.25\\ 
2007$+$776 & 110 &  23 & 1.71  &  1.41 &  2.35  & 0.045(2)  &   7.5 &  365  & 0.037(1)  &  17.5  & 0.093 & 0.151 &  0.20 &  0.39 &  3.01 & 14.10\\ 
2008$-$068 &  36 & -21 & 1.91  &  2.21 &  0.78  & 0.011(0)  &   5.6 &  365  & 0.019(0)  &  11.8  & 0.009 & 0.045 &   -   & -0.80 &  0.00 &  0.00\\ 
2032$+$107 &  55 & -17 & 2.59  &  0.84 &  0.54  & 0.021(2)  &   7.6 &  200  & 0.025(1)  &   7.5  & 0.088 & 0.178 &  0.14 & -0.34 &  0.00 &  0.00\\ 
2037$+$511 &  89 &   6 & 4.59  &  4.93 &  4.09  & 0.016(2)  &  27.5 &  365  & 0.024(1)  &   8.8  & 0.029 & 0.072 & -0.12 & -0.14 &  0.00 &  0.00\\ 
2047$+$098 &  57 & -21 & 2.19  &  0.49 &  0.64  & 0.031(2)  &   5.8 &  365  & 0.018(0)  &  13.5  & 0.046 & 0.150 &   -   &  0.21 &  2.03 &  1.66\\ 
2059$+$034 &  53 & -27 & 1.66  &  0.68 &  0.96  & 0.054(1)  &  11.7 &  365  & 0.021(0)  &   6.8  & 0.062 & 0.212 &   -   &  0.27 &  0.00 &  0.00\\ 
2105$+$420 &  85 &  -3 & 5.27  &  1.48 &  0.75  & 0.006(0)  &  18.0 &  365  & 0.014(0)  &  18.9  & 0.009 & 0.042 &   -   & -0.52 &  0.78 &  0.72\\ 
2113$+$293 &  77 & -13 & 3.26  &  0.85 &  0.87  & 0.049(2)  &   7.1 &  365  & 0.027(0)  &   5.3  & 0.126 & 0.178 &   -   &  0.02 &  0.00 &  0.00\\ 
2121$+$053 &  58 & -30 & 1.50  &  2.26 &  1.72  & 0.047(2)  &   8.5 &  365  & 0.038(0)  &  27.4  & 0.088 & 0.114 &   -   & -0.21 &  0.84 &  0.85\\ 
2134$+$004 &  55 & -36 & 1.24  &  6.94 &  6.95  & 0.019(2)  &   9.4 &  365  & 0.014(0)  &  12.3  & 0.009 & 0.029 &   -   &  0.00 &  0.95 &  0.94\\ 
2155$-$152 &  41 & -48 & 0.88  &  2.70 &  2.00  & 0.022(1)  &  12.4 &  365  & 0.019(0)  &   5.3  & 0.074 & 0.066 &   -   & -0.23 &  8.64 &  1.60\\ 
2200$+$420 &  92 & -10 & 3.43  &  3.08 &  3.40  & 0.031(2)  &   7.8 &  100  & 0.022(2)  &   4.7  & 0.225 & 0.309 &  0.66 &  0.09 &  2.66 &  1.64\\ 
2209$+$081 &  70 & -37 & 1.21  &  0.82 &  0.24  & 0.011(1)  &   9.9 &  365  & 0.024(0)  &  11.7  & 0.013 & 0.097 &   -   & -0.93 &  3.27 &  0.71\\ 
2214$+$350 &  90 & -18 & 2.39  &  0.47 &  0.57  & 0.085(2)  &   6.5 &  365  & 0.029(1)  &   6.6  & 0.087 & 0.076 &  0.48 &  0.14 &  0.00 &  0.00\\ 
2234$+$282 &  90 & -26 & 1.72  &  1.69 &  1.64  & 0.041(2)  &   7.9 &  365  & 0.021(0)  &   9.0  & 0.088 & 0.103 &   -   &  0.00 &  0.00 &  0.00\\ 
2251$+$158 &  86 & -38 & 1.18  & 12.45 & 10.04  & 0.013(1)  &   8.0 &  365  & 0.019(0)  &  13.8  & 0.067 & 0.169 &   -   & -0.09 &  1.68 &  1.07\\ 
2251$+$244 &  91 & -31 & 1.43  &  1.49 &  0.53  & 0.008(1)  &   5.4 &  365  & 0.000(0)  &   7.9  & 0.005 & 0.031 &   -   & -0.80 &  0.00 &  0.00\\ 
2307$+$107 &  87 & -45 & 1.00  &  0.37 &  0.36  & 0.025(1)  &   3.1 &  365  & 0.022(0)  &  15.7  & 0.052 & 0.130 &   -   & -0.02 &  0.00 &  0.00\\ 
2319$+$272 & 100 & -31 & 1.37  &  0.80 &  0.58  & 0.017(1)  &   5.1 &  365  & 0.015(0)  &  29.2  & 0.033 & 0.061 &   -   & -0.24 &  0.00 &  0.00\\ 
2344$+$092 &  97 & -50 & 0.87  &  1.68 &  1.19  & 0.008(1)  &   7.3 &  365  & 0.013(0)  &  14.7  & 0.012 & 0.059 &   -   & -0.26 &  0.76 &  1.78\\ 
2352$+$495 & 114 & -12 & 2.58  &  2.18 &  1.09  & 0.016(2)  &   9.2 &  365  & 0.016(0)  &   7.3  & 0.010 & 0.016 &   -   & -0.49 &  1.02 &  0.89\\  
\enddata 
\label{tab:results}
\end{deluxetable}

\end{document}